\documentclass[prd,showpacs,superscriptaddress]{revtex4}
\usepackage{amsmath}
\usepackage{amssymb}
\usepackage{color}
\usepackage{graphicx}

\begin{document}

\title{The spectral shapes of the fluxes of electrons and positrons \\
and the average residence time of cosmic rays in the Galaxy} 

\author{Paolo Lipari}
\email{paolo.lipari@roma1.infn.it}
\affiliation{INFN, Sezione Roma ``Sapienza'',
 piazzale A.Moro 2, 00185 Roma, Italy}


\begin{abstract}
The cosmic ray energy spectra encode
very important information about the mechanisms that generate relativistic
particles in the Milky Way, and about the properties of the Galaxy that control
their propagation. 
Relativistic electrons and positrons traveling in interstellar space lose energy
much more rapidly than more massive particles such as protons and nuclei,
with a rate that grows quadratically with the particle energy $E$.
One therefore expects that the effects of energy loss should
leave observable signatures in the $e^\mp$ spectra,
in the form of softenings centered at the critical energy $E^*$.
This quantity is determined by the condition that
the total energy loss suffered by particles during
their residence time in the Galaxy is of the same order of the initial energy.
If the electrons and positrons are accelerated in
discrete (quasi) point--like astrophysical objects,
such as Supernova explosions or Pulsars,
the stochastic nature of the sources should also leave
observable signatures in the $e^\mp$ energy spectra and angular distributions
above a second (higher) critical energy $E^\dagger$, determined by the condition that
particles with $E \gtrsim E^\dagger$ can propagate only for a
maximum distance shorter than the average separation between sources.
In this work we discuss the theoretically expectations
for the signatures of the energy loss effects on the electron
and positron spectra, and compare these predictions with the
existing observations.
Recent measurements of the $(e^- + e^+$) flux have
discovered the existence of a prominent spectral break at $E \simeq 1$~TeV.
This spectral feature can perhaps be identified with the critical
energy $E^*$. An alternative hypothesis is
to assmume that $E^*$ has a much lower value of order few GeV. 
Resolving this ambiguity is
of great importance for our understanding of Galactic cosmic rays.
\end{abstract}

\maketitle

\section{Introduction}
\label{sec:intro} 
The fluxes of cosmic rays (CR) that are observable near the Earth,
in a very broad energy interval that extends from GeV's to PeV's
(and possibly to much higher energy) are formed by particles of Galactic origin.
These CR particles are injected in interstellar space by sources distributed in
the entire Milky Way, and remain confined in the Galaxy for an extended
period of time because of the existence of the Galactic magnetic fields.
The CR fluxes are in very good approximation isotropic,
but their energy distributions contain very important information
about their sources and about the properties of the Galaxy
that control CR propagation. 
At low energy ($E \lesssim 30$~GeV) the CR spectra are
also distorted by the electromagnetic fields that fill the heliosphere,
generating time dependent solar modulations.

A fundamental problem for CR astrophysics is to disentangle the roles
of the sources and of propagation in the formation of the CR fluxes,
reconstructing from the observations the ``source spectra''
(that is the spectra with which the particles are injected in interstellar space
by the sources), and the distortion effects generated by propagation.

Relativistic, electrically charged particles propagating in
interstellar space lose continuosly energy because of
different types of interactions.
The rate of energy loss (for particle of the same $E$)
is however many order of magnitude larger for particles
of small mass (with the dominant mechanisms scaling $\propto m^{-4}$).
It is therefore expected that
the energy loss effects play a significant role
only in the propagation of electrons and positrons,
and be entirely negligible in the propagation of
$p$, $\overline{p}$ and nuclei.

The rate of energy loss for $e^\mp$ grows also rapidly 
with the particle energy (in good approximation $\propto E^{2}$), so
the effects of energy lossses are significant only
for particles of sufficiently high $E$.
The conclusion is that the spectra of
both the $e^-$ and $e^+$ fluxes should exhibit the ``imprint'' of
the energy losses in the form of a softening feature that marks
the transition from a regime where energy losses are negligible
($E \lesssim E^*$)
to a regime where energy losses are the most important 
``sink'' that balances the injection of new particles by the sources
($E \gtrsim E^*$).
The identification of the softening features
in the $e^\mp$ spectra allows a measurement of the critical energy $E^*$.

The physical condition that determines $E^*$ is that
the total energy lost by a particle injected in interstellar space
with initial energy $E^*$, calculated integrating
the loss rate during the time
interval when the particle remains confined in the Galaxy, 
is (on average) of the same order of magnitude.
This condition can be expressed in the form:
\begin{equation}
 \left \langle \Delta E (E^*) \right \rangle =
\langle dE/dt (E^*) \rangle ~T_{\rm esc}(E^*) \approx E^* ~.
\label{eq:estar0}
\end{equation}
In this equation $\langle dE/dt (E)\rangle$ is the rate of energy loss
averaged over the CR confinement volume and $T_{\rm esc}(E)$ is the average
residence time of a particle of initial energy $E$.
The crucially important point is that the identification
of the critical energy $E^*$ allows a measurement of the residence time
of CR in the Galaxy, a quantity of fundamental importance for
our understanding of cosmic rays. 

In the discussion above we have implicitly assumed that the critical
energies for electrons an positrons $E^*_{e^-}$ and $E^*_{e^+}$
are approximately equal.
This is a simple and robust prediction that can be used
to test the hypothesis that a spectral feature
observed in the spectrum of one particle type 
is indeed generated by energy loss effects.
If this is the case a softening feature of approximately the same
structure, and centered at approximately the same energy must 
also be present in the spectrum of the other particle type.
It should however be noted that the
two critical energies for electrons and positrons
are not exactly identical.
The rates of energy loss for $e^+$ and $e^-$ in interstellar space
are in very good approximation equal,
and the annihilation probability for CR positrons
interacting with the electrons of the medium is negligibly small,
however, the critical energy 
also depends on the space distributions of the
CR sources, that can be different for the two particle types.
In addition one should take into account of the fact that the
spectral distortion generated by the energy losses depend on the
shape of the source spectrum, that again is different
for the two particle types.

When energy losses become important for the propagation
of electrons and positrons, the space--time volume that contains
sources that can contribute to the observed spectra starts to shrink
rapidly for increasing $E$. Ìf the CR sources are
discrete and transient astrophysical objects, the reduction
in the space--time volume for the CR sources
corresponds to a smaller number of astrophysical objects 
that can contribute to the flux. When this number becomes sufficiently small,
the ``granularity'' (or ``stochasticity'')
of the sources should become observable as an anisotropy
in the angular distribution of the flux and/or in additional features in
the energy spectra. The search for these predicted source granularity
effects is an important task in the study of the $e^\mp$ spectra.

This work is organized as follows:
in the next section we discuss the observations of the $e^-$, $e^+$ and
($e^- + e^+$) spectra, fitting the data with simple functional forms.
Section~\ref{sec:models} discusses theoretical
models for the formation of the Galactic CR spectra.
In this discussion we introduce two very simple models
for Galactic propagation, where the CR propagation can be expressed
with exact analytic expressions. The first one is the ``Leaky Box'' model,
the second one is a diffusion model that gives results that are
very close to those that can be obtained numerically with computer codes
such as GALPROP or DRAGON. These models should be considered as only
first order approximations to a realistic description of CR propagation,
but can be used as a qualititative (or semi--quantitative) guide to the
expected spectral shapes for different particle types.
The diffusion model also allows to study the size and shape of
the space region where the CR particles have been produced.
In section~\ref{sec:discrete} we assume that CR are generated in
``source events'' (such as supernovae explosions)
that are point--like and instantaneous and, using the framework 
of the diffusion model, compute how the observed flux is formed
by the contributions of different source events, and how
the granularity of the source could become observable.
Section~\ref{sec:interpretation} discusses how the theoretical models
compare to the observations. We argue that there are two
possible solutions for the critical energy $E^*$ ($E^* \sim 3$~GeV)
or $E^* \simeq 1$~TeV. These two solutions have profoundly different
implications for cosmic ray astrophysics, as discussed in
Sec.~\ref{sec:astro}. A final section gives a summary and an outlook for
future studies.

\section {Observations} 
\label{sec:observations}
\begin{figure}[bt]
\begin{center}
\includegraphics[width=11.0cm]{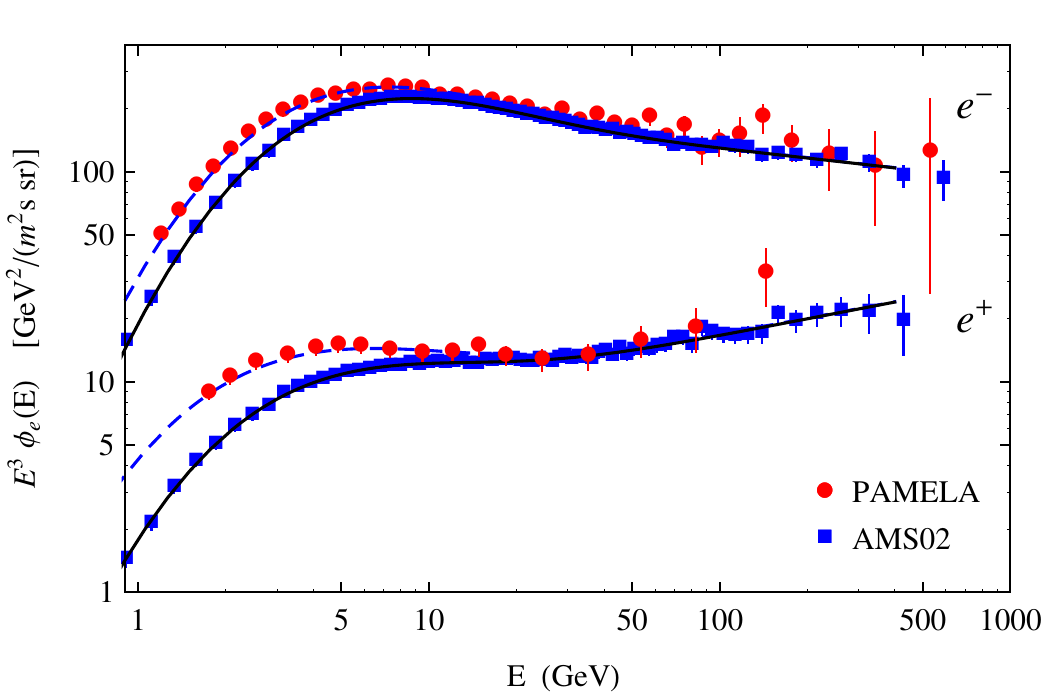}
\end{center}
\caption {\footnotesize
Spectra of $e^\mp$ measured
by PAMELA \cite{pamela-electrons,pamela-positrons} (squares)
and AMS02 \cite{ams02-electrons-positrons-2014} (circles)
shown in the form $E^3 \, \phi_{e^\mp} (E)$ versus $E$.
The energy scale of the PAMELA has been rescaled by a factor $f = 0.93$.
The lines are fit to the data discussed in the text. 
\label{fig:spectra-magnetic}}
\end{figure}

\begin{figure}[bt]
\begin{center}
\includegraphics[width=11.0cm]{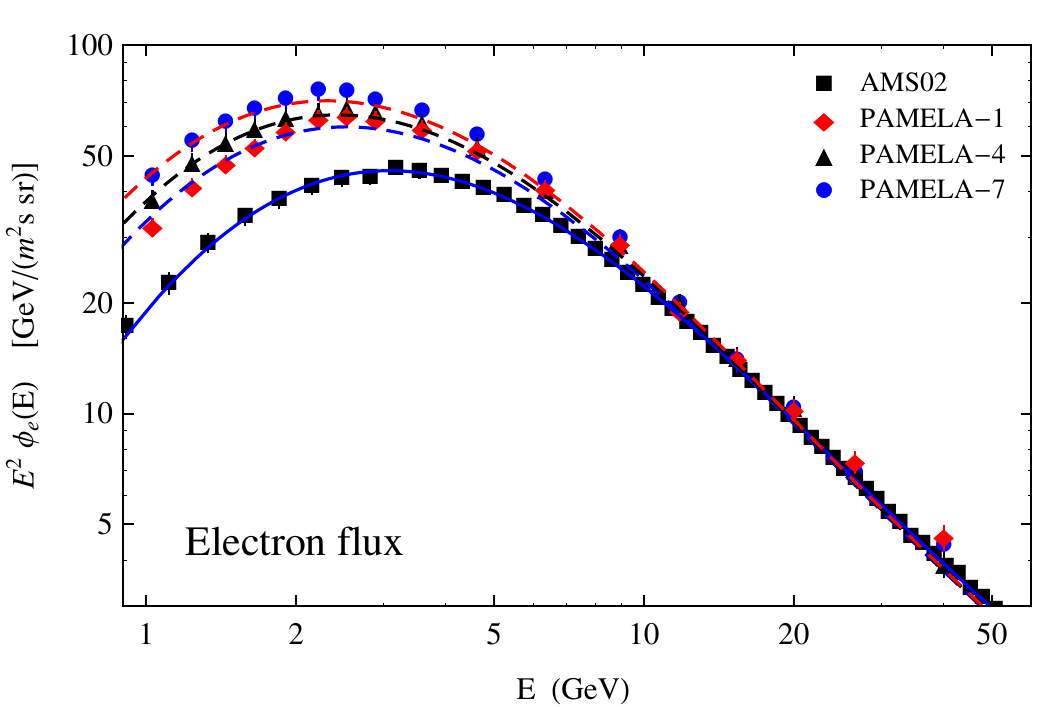}
\end{center}
\caption {\footnotesize
Electron spectra measured by AMS02 \cite{ams02-electrons-positrons-2014}
and by PAMELA \cite{Adriani:2015kxa} in different time intervals.
The energy scale of PAMELA has been rescaled by a factor $f = 0.93$.
The lines are fit to the data discussed in the main text. 
\label{fig:spectra-magnetic1}}
\end{figure}

The spectra of cosmic ray (CR) electrons and positrons have been measured
with good accuracy by the magnetic spectrometers
PAMELA \cite{pamela-electrons,pamela-positrons,Adriani:2014pza,Adriani:2015kxa,Adriani:2016uhu}
and AMS02 \cite{ams02-electrons-positrons-2014,Aguilar:2018ons} for $E \lesssim 500$~GeV
(see Figs.~\ref{fig:spectra-magnetic} and~\ref{fig:spectra-magnetic1}).
The two detectors have taken data during non overlapping time intervals,
and the large differences in flux observed for $E \lesssim 20$~GeV,
can be attributed to time variations associated to the 
effects of (time dependent) solar modulations that distort a (constant) interstellar spectrum.

In the discussion of the CR energy distributions it is useful to
to consider the (energy dependent) spectral index $\gamma(E)$,
that is the slope of the spectrum
when shown in the form $\log\phi$ versus $\log E$: 
\begin{equation}
\gamma (E) = - \frac{d\log \phi}{d\log E} = - \frac{E}{\phi} \; \frac{d\phi}{dE} ~.
\label{eq:spectral-index}
\end{equation}

The energy spectra of both electrons and positrons have the following qualitative properties.
\begin{itemize}
\item [(i)] For $E \lesssim 20$~GeV 
(that is also the region where time dependent solar modulations are important)
the spectra exhibit a strong curvature and soften gradually, as
the spectral indices $\gamma_{e^\mp} (E)$ grow continuously with energy.
\item [(ii)] At high energy ($E \gtrsim 50$~GeV) the
 spectra can be well fitted with simple power laws, and
 the spectral indices are approximately constant
 (with values 
 $\gamma_{e^-} \simeq 3.17$ and
 $\gamma_{e^+} \simeq 2.74$).
\item [(iii)] The spectral indices of both electrons and positrons
 have a maximum at an energy of order 20~GeV (more accurately $E \simeq 24$~GeV for electrons and
 $E \simeq 14$~GeV for positrons).
 This imply that both spectra undergo a modest, but clearly
 visible hardening for energy around 10--30~GeV.
\end{itemize}

To describe the data with have used a simple model where the
interstellar spectra have the a broken power law form
(with one hardening), and the solar modulations are described by the
Force Field approximation \cite{Gleeson:1968zza}.
A convenient expression
to descrive a spectrum with one
gradual break
(see \cite{Lipari:2017jou} for a more detailed discussion) is:
\begin{equation}
 \phi_0 (E) = K_0 \;
 \left ( \frac{E}{E_0} \right )^{-\gamma_1} \;
 \left [1 + \left (\frac{E}{E_b} \right )^{\frac{1}{w}} \right ]^{-(\gamma_2 - \gamma_1) \, w} ~.
\label{eq:break-parametrization} 
\end{equation}
This 5--parameters expression
(with $E_0$ an arbitrary reference energy that in this work will be fixed at 10~GeV)
describes a spectrum that is asymptotically (at low and high energy)
a simple power law with slopes $\gamma_1$ and $\gamma_2$.
At the break energy $E_b$ the spectral index takes the average value
$\gamma (E_b) = (\gamma_1 + \gamma_2)/2$, and $w$ gives the width of the interval in
$\log E$ where spectral index varies.

In the Force Field Approximation (FFA) \cite{Gleeson:1968zza}
the effects of solar modulations are modeled as the
loss of a constant energy $\Delta E = \varepsilon$ for all particles
that traverse the heliosphere and reach the Earth.
Making use of the Liouville theorem, 
and assuming that the CR flux is isotropic at the boundary of the heliosphere,
one can derive the relation between the spectrum $\phi (E)$ observed at the Earth
and the interstellar spectrum $\phi_0 (E)$ (present at the boundary of the heliosphere):
\begin{equation}
 \phi (E) = \frac{p^2}{p_{0}^2} \; \phi_{0} (E+\varepsilon)
\end{equation}
where $p$ and $p_{0}$ are the momenta that correspond to the energies
$E$ and $E_{0} = E + \varepsilon$.

\begin{figure}[bt]
\begin{center}
\includegraphics[width=7.0cm]{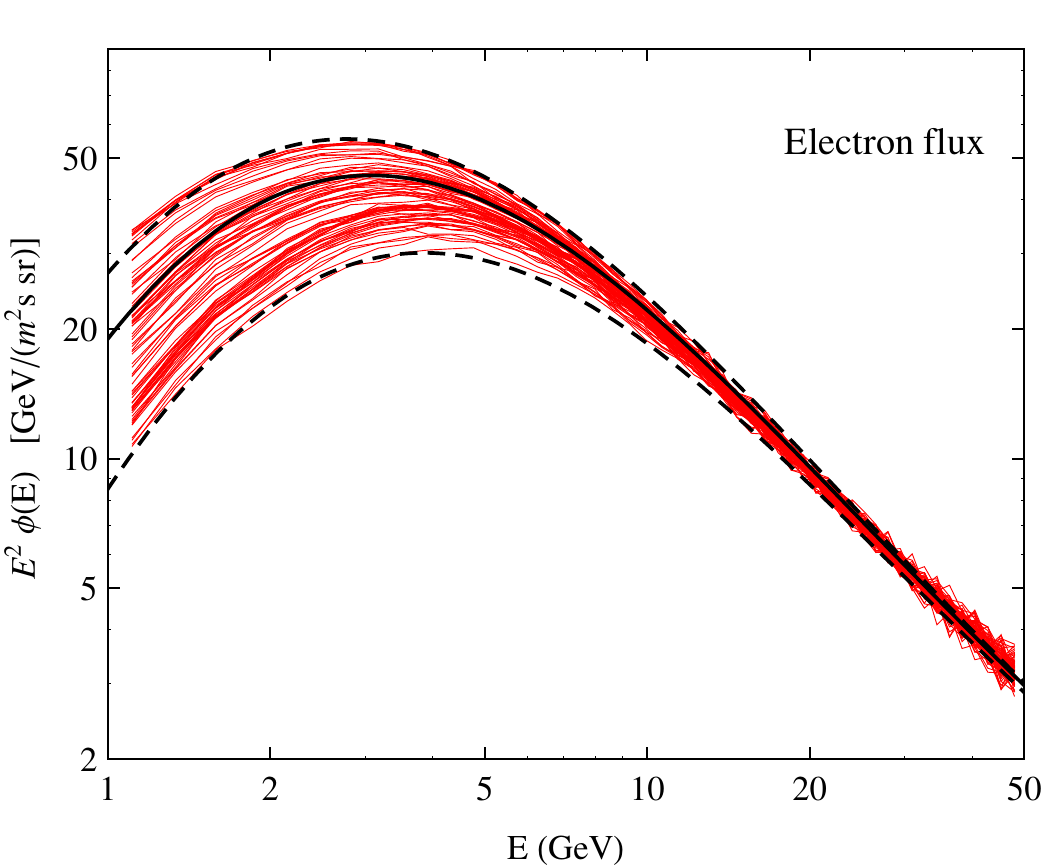}
~~~~
\includegraphics[width=7.0cm]{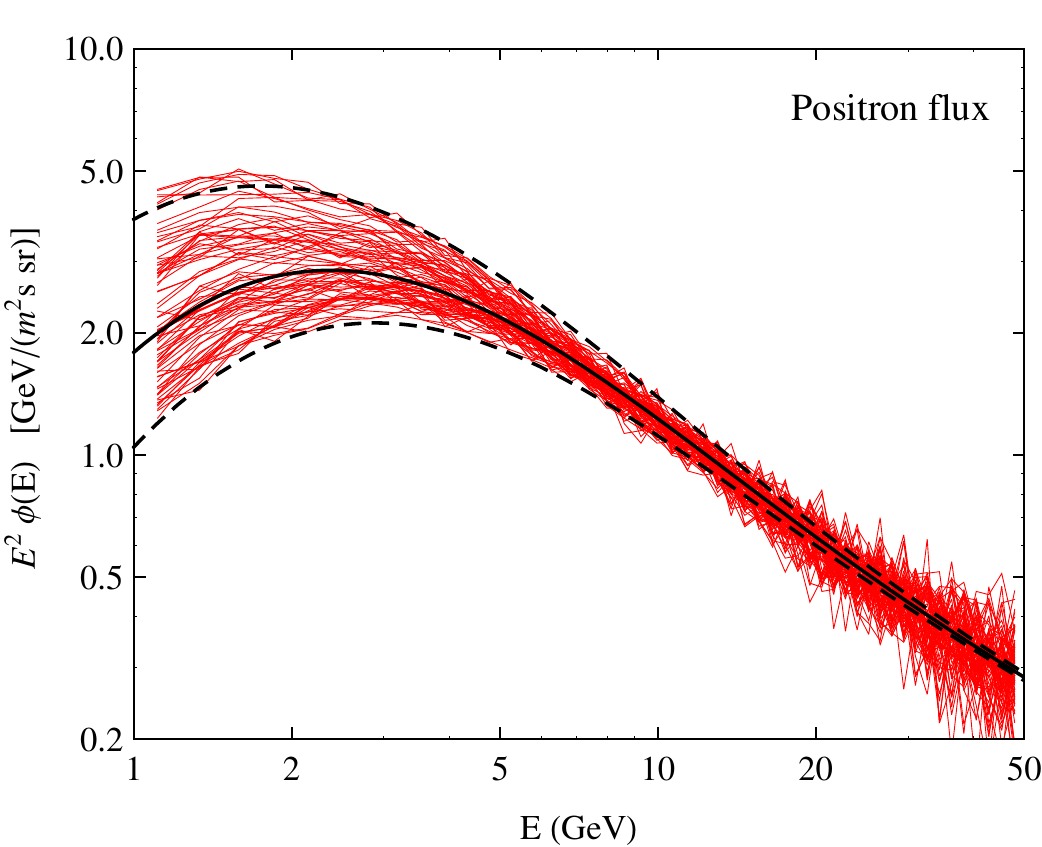}
\end{center}
\caption {\footnotesize
 Measurements of the electron (left panel) and positron (right panel) spectra
 taken during 79 different time intervals (of 27 days) by the AMS02
 collaborations \protect\cite{Aguilar:2018ons}.
 The thick solid lines are
 the fits of the timne averaged $e^\mp$ spectra discussed in this work.
 The dashed lines are calculated distorting the average spectra using
 the FFA model for solar modulations with different parameters.
\label{fig:ele-time}}
\end{figure}

It is now firmly established that the solar modulation effects for positively
and negatively charged particles are different. This is a simple
consequence of the large scale heliospheric magnetic field. Particles 
with electric charge of different sign travel along 
different trajectories and therefore on average can lose different
amount of energy. Accordingly, one expects that the parameter $\varepsilon$ will be
different for electrons and positrons.

Applying the FFA algorithm to a spectrum of the form
of Eq.~(\ref{eq:break-parametrization}),
and considering ultrarelativistic electrons (so that $p \simeq E$ and
$p_{0} \simeq E_{0}$), one arrives to the (six parameter) expression:
\begin{equation}
\phi(E) = K_0 ~
 \frac{E^2}{(E+ \varepsilon)^2} ~
 \left ( \frac{E+ \varepsilon}{E_0} \right )^{-\gamma_1} 
 \left [ 1 +
 \left (\frac{E+ \varepsilon}{E_b} \right )^{1/w} \right ]^{-(\gamma_2 - \gamma_1) \, w} 
~.
 \label{eq:fit-form1}
\end{equation}

The expression of Eq.~(\ref{eq:fit-form1}) can describe well the
AMS02 data (taken between May 2011 and November 2013)
for both electrons and positrons in the energy range $E > 1$~GeV. 
The minimum $\chi^2$ (calculated combining 
quadratically statistical and systematic errors)
are $\chi^2_{\rm min} = 11.7$ (64 d.o.f.)
for the electrons, and $\chi^2_{\rm min} = 21.0$ (63 d.o.f.) for the positrons. 
The results are shown in Fig.~\ref{fig:spectra-magnetic}, and the best fits parameters,
with 1~$\sigma$ uncertainties, are listed in Table~\ref{tab:fit1}.

\begin{table}
 \caption{\footnotesize
 Parameters of best fits to the AMS02
 data on the electron and positron spectra
 \cite{ams02-electrons-positrons-2014} (in the range $E > 1$~GeV),
 using the functional form of Eq.~(\ref{eq:fit-form1}).
 The $\chi^2$ of the fit is calculated combining quadratically the stated
 statistical and systematic errors. 
 The best fit spectra are shown in Fig.~\ref{fig:spectra-magnetic}.
 \label{tab:fit1}}
\begin{center}
 \renewcommand{\arraystretch}{1.65}
 \begin{tabular}{ | l || c | c |}
\hline
Particle & Electrons & Positrons \\
\hline 
$K$ [GeV\;m$^2$\;s\;sr]$^{-1}$ & $0.47^{+0.02}_{-0.03}$ & $0.018^{+0.0012}_{-0.003}$ \\
$\gamma_1$ & $3.89^{+0.27}_{-0.12}$ & $3.62^{+0.36}_{-0.22}$ \\
$\gamma_2$ & $3.17^{+0.05}_{-0.09}$ & $2.74^{+0.05}_{-0.15}$ \\
$E_b$ [GeV] & $32.9^{+5.3}_{-7.2}$ & $20.8^{+5.3}_{-12.4}$ \\
$w$ & $0.50^{+0.25}_{-0.15}$ & $0.55^{+0.29}_{-0.21}$ \\
$\varepsilon$ [GeV] & $1.44^{+0.31}_{-0.10}$ & $0.94^{+0.36}_{-0.14}$ \\
\hline
$\chi^2_{\rm min}$ & 11.7 & 21.0 \\
$N_{\rm d.o.f.}$ & $64$ & $63$ \\
\hline
\end{tabular}
\end{center}
\end{table}

The PAMELA collaboration has published spectra of
electrons \cite{pamela-electrons} and positrons \cite{pamela-positrons},
taken in the time interval from January 2006 to December 2009.
In the framework we have adopted here, the data taken
in different time interval should be fitted 
with the same expression used for AMS02, only
changing the (time dependent) parameter $\varepsilon$ in Eq.~(\ref{eq:fit-form1}).
We will however also allow for the possibility that the energy
scales of the two experiments are different.
Following these ideas we have studied the hypothesis 
that the PAMELA measurements of the $e^\mp$ spectra can be described
by the same best fit functions $\phi_{e^\mp}^{\rm AMS} (E)$,
with a rescaling of the energy and the distortion
due to a difference in the parameter $\varepsilon$.

This correspond to fit the PAMELA data from \cite{pamela-electrons,pamela-positrons}
with an expression that depends on two parameters.
One parameter is the ratio between the scale of energy for
the two detectors $f = E_{\rm AMS}/E_{\rm PAM}$. The second one
is the quantity $\Delta \varepsilon = \varepsilon_{\rm PAM} - \varepsilon_{\rm AMS}$
that is the difference between the solar modulations parameters
for the two data taking time intervals.

This procedure yields best fits of reasonable good quality
(in the energy range $E > 1$~GeV).
For the electron spectrum one has $\chi^2_{\rm min} = 26.2$ (for 37 d.o.f.),
while for the positron flux one has
$\chi^2_{\rm min} = 4.8$ for 15~d.o.f..

The parameters of the best fit are given in
table~\ref{tab:fit2}. 
The best fit value for the parameter $\Delta \varepsilon$ is
$-210 \pm 10$~MeV for electrons and
$-280 \pm 20$~MeV for positrons. 
The best fit values for the parameters 
$f_{e^\mp}$ show significant deviations from unity
($f_{e^-} \simeq 0.93\pm 0.01$ and $f_{e^+} \simeq 0.94\pm 0.02$).
It is encouraging that these two values are
consistent with being equal to each other.
It should also be noted that a rewiew of the PAMELA data
\cite{Adriani:2014pza} discusses a revision of the normalization
of the published data, in the same direction we find in the fit procedure.

\begin{table}
 \caption{\footnotesize
Parameters of fits to the PAMELA data on
the electron \cite{pamela-electrons} 
and positron \cite{pamela-positrons} fluxes.
The data is fitted using the best fit function
obtained for the AMS data, assuming a constant energy rescaling factor
$f = E_{\rm PAM} /E_{\rm AMS}$, and a difference in the FFA modulation
parameter $\Delta \varepsilon = \varepsilon_{\rm PAM} - \varepsilon_{\rm AMS}$.
 \label{tab:fit2}}
\begin{center}
 \renewcommand{\arraystretch}{1.25}
 \begin{tabular}{ | l || c | c |}
\hline
Particle & Electrons & Positrons \\
\hline 
$f$ & $0.93 \pm 0.01$ & $0.94 \pm 0.02$ \\
$\Delta \varepsilon$ [GeV] & $-0.21 \pm 0.01$ & $-0.28 \pm 0.02$ \\
\hline
$\chi^2_{\rm min}$ & 26.2 & 4.8 \\
$N_{\rm d.o.f.}$ & $37$ & $15$ \\
\hline
\end{tabular}
\end{center}
\end{table}

These fits to the PAMELA data are shown in Fig.~\ref{fig:spectra-magnetic}.
[Note that in the figure the PAMELA data points (plotted together with the AMS02 data)
have the energy rescaled by a factor 0.93.]

The PAMELA collaboration has also published
the results of seven measurements of the electron flux 
taken during separate six--months time intervals
between July 2006 and December 2009 \cite{Adriani:2015kxa}.
These results are limited to the lower energy region ($E \lesssim 50$~GeV),
but are sufficient to obtain important information on the time dependence of the $e^-$ flux.
These spectra have also been successfully
fitted ($\chi^2_{\rm min}$ is in the range 1.0--8.3 for 14 d.o.f).
with the 2--parameter model discussed above.
The fits are shown in Fig.~\ref{fig:spectra-magnetic1},
and the best fit for the parameters $\{f, \Delta \varepsilon\}$ are listed in Table~\ref{tab:fit3}.
The energy rescale parameter in the different fits is consistent with being equal
for all data sets $f \simeq 0.97$. The solar modulation dependent parameter
$\Delta \varepsilon$ takes values between $-200$ and $-310$~MeV.

\begin{table}
 \caption{\footnotesize
Parameters of fits to the electron PAMELA
taken during different time intervals \cite{Adriani:2015kxa}.
The data is fitted using the best fit function
obtained for the AMS data, assuming a constant energy rescaling factor
$f = E_{\rm PAM} /E_{\rm AMS}$, and a difference in the FFA modulation
parameter $\Delta \varepsilon = \varepsilon_{\rm PAM} - \varepsilon_{\rm AMS}$.
 \label{tab:fit3}}
\begin{center}
 \renewcommand{\arraystretch}{1.25}
 \begin{tabular}{ | l || c | c | c| c | }
\hline
Data set & Time & $f$ & $\Delta \varepsilon$ & $\chi^2_{\rm min}$ \\
\hline 
1 & Jul--Nov 2006 & $0.975 \pm 0.013$ & $-0.208 \pm 0.013$ & 7.7 \\
2 & Jan--Jun 2007 & $0.974 \pm 0.012$ & $-0.243 \pm 0.012$ & 3.9 \\
3 & Jul--Dec 2007 & $0.967 \pm 0.012$ & $-0.256 \pm 0.012$ & 8.3 \\
4 & Jan--Jun 2008 & $0.960 \pm 0.013$ & $-0.259 \pm 0.012$ & 1.0 \\
5 & Jul--Dec 2008 & $0.962 \pm 0.014$ & $-0.284 \pm 0.012$ & 2.7 \\
6 & Jan--Jun 2009 & $0.956 \pm 0.014$ & $-0.310 \pm 0.013$ & 3.2 \\
7 & Jul--Dec 2009 & $0.976 \pm 0.018$ & $-0.307 \pm 0.014$ & 5.6 \\
\hline
\end{tabular}
\end{center}
\end{table}

More recently the AMS02 collaboration has published \cite{Aguilar:2018ons}
spectra of electrons and positrons in the energy range 1--50~GeV
averaged during 79 time separate time intervals of 27 days.
The FFA model for the solar modulations
cannot reproduce exactly these precisely measured spectra 
however in a reasonably good first approximation it still gives a good description of the data,
as shown in Fig.~\ref{fig:ele-time}.
A more detailed study of this problem is postponed to a future work.

\subsection{The ($e^- + e^+$) spectrum}
\begin{figure}[bt]
\begin{center}
\includegraphics[width=11.0cm]{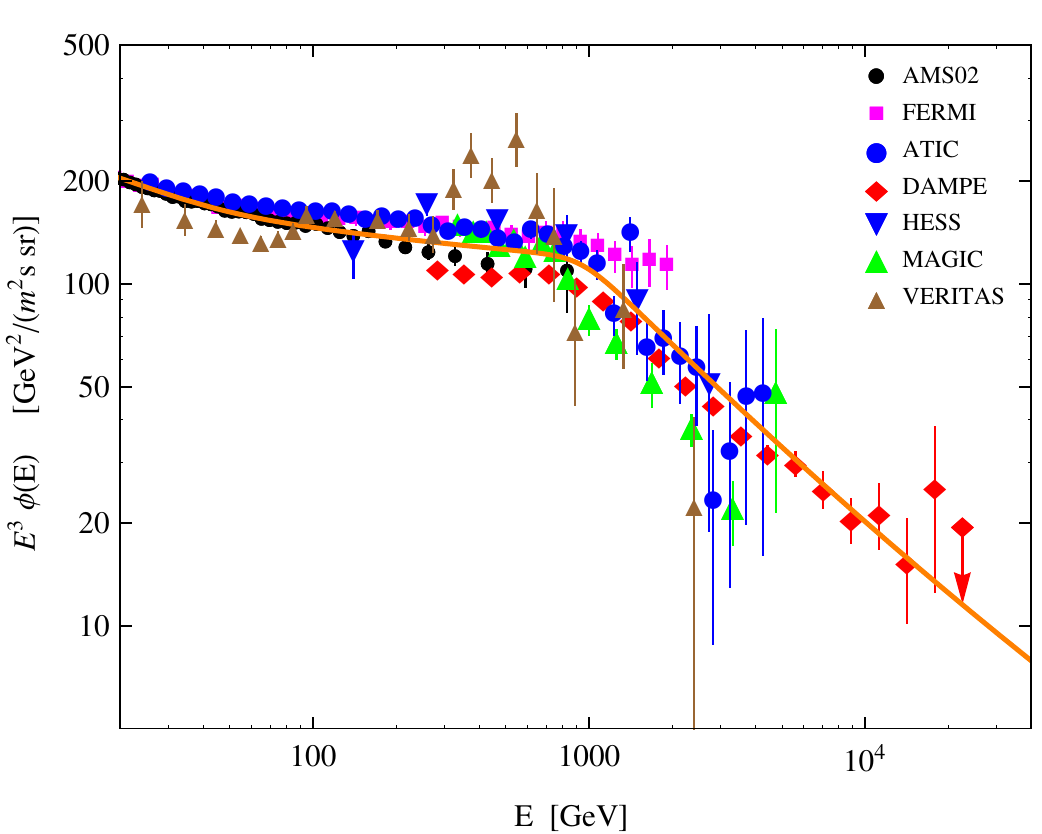}
\end{center}
\caption {\footnotesize
Measurements of the ($e^- + e^+)$ spectrum obtained by
AMS02 \protect\cite{ams02-allelectrons-2014},
FERMI \protect\cite{Abdollahi:2017nat},
ATIC \protect\cite{Chang:2008aa},
DAMPE \protect\cite{Ambrosi:2017wek},
HESS \protect\cite{Aharonian:2008aa,Aharonian:2009ah,hess-icrc2017},
MAGIC \protect\cite{BorlaTridon:2011dk},
and VERITAS \cite{Staszak:2015kza}.
The line is a fit to the data from
\protect\cite{Lipari:2018gzn}
\label{fig:allele1}}
\end{figure}

\begin{figure}[bt]
\begin{center}
\includegraphics[width=11.0cm]{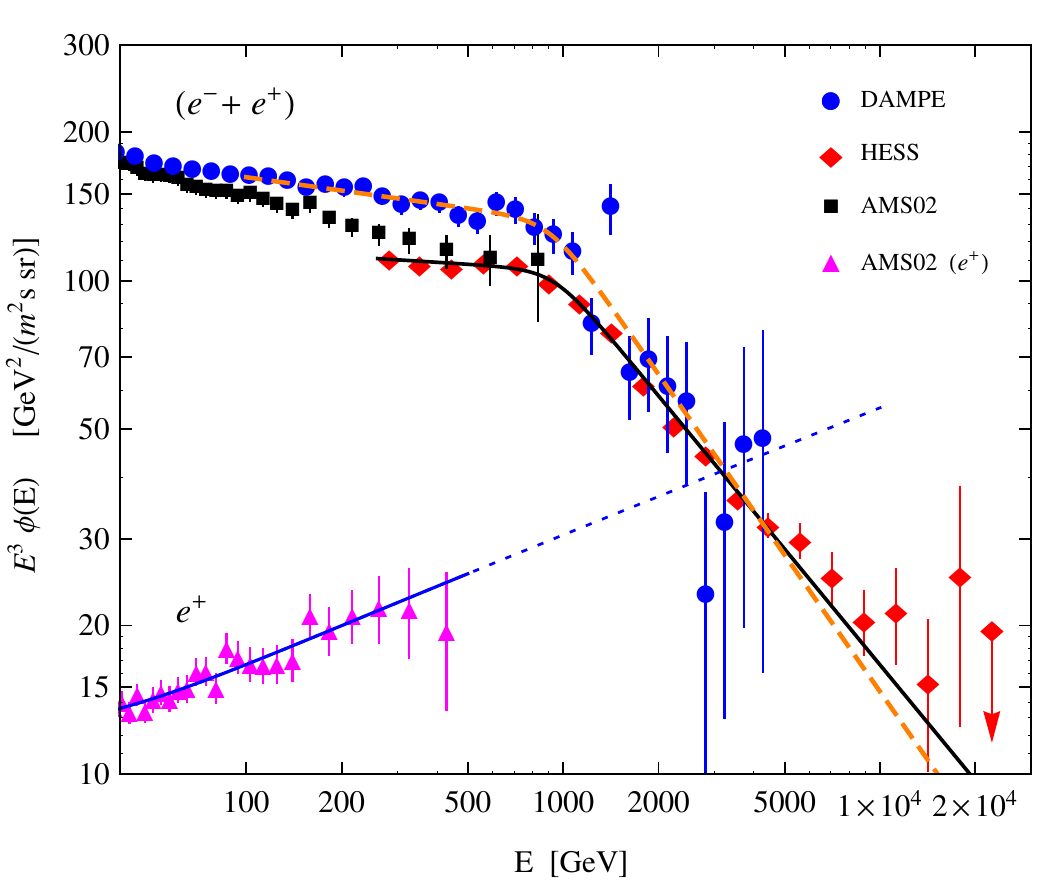}
\end{center}
\caption {\footnotesize
Measurements of the ($e^- + e^+)$ spectrum obtained by
AMS02 \protect\cite{ams02-allelectrons-2014},
DAMPE \protect\cite{Ambrosi:2017wek}
and HESS,
and of the $e^+$ flux by AMS02.
The lines are fits of the DAMPE \cite{Ambrosi:2017wek}
and HESS data \cite{hess-icrc2017} data.
\label{fig:allele2}}
\end{figure}

Calorimetric measurements of the sum of electrons and positrons,
performed without separating the two particle types
are shown in Fig.~\ref{fig:allele1} and~\ref{fig:allele2}.
Direct observations of this spectrum 
have been performed by AMS02 \cite{ams02-allelectrons-2014},
FERMI \cite{Abdollahi:2017nat},
ATIC \cite{Chang:2008aa} and DAMPE \cite{Ambrosi:2017wek}.
These measurements reach (in the case of DAMPE) a maximum energy $E \simeq 4.5$~TeV.

Ground based Cherenkov telescopes
such as HESS \cite{Aharonian:2008aa,Aharonian:2009ah}, 
MAGIC \cite{BorlaTridon:2011dk} and VERITAS \cite{Staszak:2015kza}, 
have also been able to obtained measurements of the $(e^- + e^+)$ spectrum in the TeV energy range.
The measurements are performed selecting events 
that are consistent with an electromagnetic shower,
and subtracting the background generated by cosmic ray
protons and nuclei using Montecarlo codes to model the developments of air showers.
Recently the HESS telescope \cite{hess-icrc2017}
has presented measurements of the spectrum that extend to $E \simeq 20$~TeV.

The observations of ATIC \cite{Chang:2008aa}
show a surprising structure in the spectrum at an energy of 700~GeV, that
has been the object of many efforts at interpretation, generating
large body of literature. The existence of this structure
has however not been confirmed by the measurements of FERMI, DAMPE, HESS and the other
Cherenkov detectors, and in the present work will be considered as
the consequence of some unaccounted for systematic effects.

The data of the other experiments are not in perfect agreement with each other,
for example (see Fig.~\ref{fig:allele2})
the measurements of DAMPE and HESS differ by approximately 25--30\% in the energy range 500--1000~GeV.
The observations however show clearly the existence of a remarkable
spectral break at $E \approx 1$~TeV.

The existence of this spectral feature
has been first discovered by the HESS Collaboration \cite{Aharonian:2009ah}, 
confirmed by MAGIC \cite{BorlaTridon:2011dk} and VERITAS \cite{Staszak:2015kza},
and then also clearly seen by DAMPE detector \cite{Ambrosi:2017wek}.
The results of FERMI \cite{Abdollahi:2017nat} 
are actually consistent with an unbroken power law spectrum, but
the errors of the highest energy points, that reach 2~TeV are large,
and this does not appear to be a significant discrepancy.

The HESS Collaboration \cite{hess-icrc2017} 
has presented a fit to their more recent data
with the same expression of Eq.~(\ref{eq:break-parametrization})
obtaining for the best fit parameters 
a break energy $E_{b} \simeq 0.94 \pm 0.02 ^{+0.29}_{-0.21}$~TeV,
spectral indices $\Gamma_1 = 3.04 \pm 0.01^{+0.010}_{-0.18}$ (below the break)
and $\Gamma_2 = 3.78 \pm 0.02^{+0.017}_{-0.06}$ (above the break), and a 
narrow width $w = 0.12 \pm 0.01^{+0.19}_{-0.05}$.

The DAMPE Collaboration fits the data using a form that is
a little different from the form of Eq.~(\ref{eq:break-parametrization})
adopted here, and used by HESS:
\begin{equation}
 \phi(E) = K \;
 \left ( \frac{E}{E_0} \right )^{-\gamma_1} \;
 \left [1 + \left (\frac{E}{E_b} \right )^{-(\gamma_2 - \gamma_1)/s} \right ]^{s} ~.
\label{eq:break-par-ams} 
\end{equation}
Equations~(\ref{eq:break-parametrization}) and~(\ref{eq:break-par-ams}) are
different parametrizations of the same ensemble of curves, but 
the parameter $s$, used in Eq.~(\ref{eq:break-par-ams}),
and the width $w$ in Eq.~(\ref{eq:break-parametrization})
are not identical \footnote{The use of the width parameter $w$
 is preferable because it has a clear and intuitive physical meaning as the width of the
 spectral break.
In this case (where $\Delta \gamma \simeq 1$) $s$ and $w$ are approximately the same,
but more in general the two quantities can be very different.},
but related by:
\begin{equation}
s = {w}/{|\Delta \alpha|} ~.
\label{eq:s-w}
\end{equation}
DAMPE finds for best fit parameters a 
break energy $E_{b} \simeq 0.914 \pm 0.098$~TeV, $\Gamma_1 = 3.09 \pm 0.01$,
$\Gamma_2 = 3.92 \pm 0.20$, with the value of $s$ fixed at 0.1
(that corresponds roughly to $w = 0.083$).

Also VERITAS have published a fit \cite{Staszak:2015kza} with a broken power law form
(that is $w \to 0$) where the best fit break energy
$E_{\rm Veritas} \simeq 0.71 \pm 0.04$~TeV, is somewhat lower than the estimates
of HESS and DAMPE. 

The not perfect agreement between the different measurements suggest the
existence of some systematic effects, and for this reason we will not
attempt to perform a global fit to all data.
Even in the presence of these uncertainties, 
the existence of a sharp break in the spectrum
of the sum ($e^- + e^+$) should be considered as solidly established.

\subsection{Observed features in the $e^\mp$ spectra}
\begin{figure}[bt]
\begin{center}
\includegraphics[width=11.0cm]{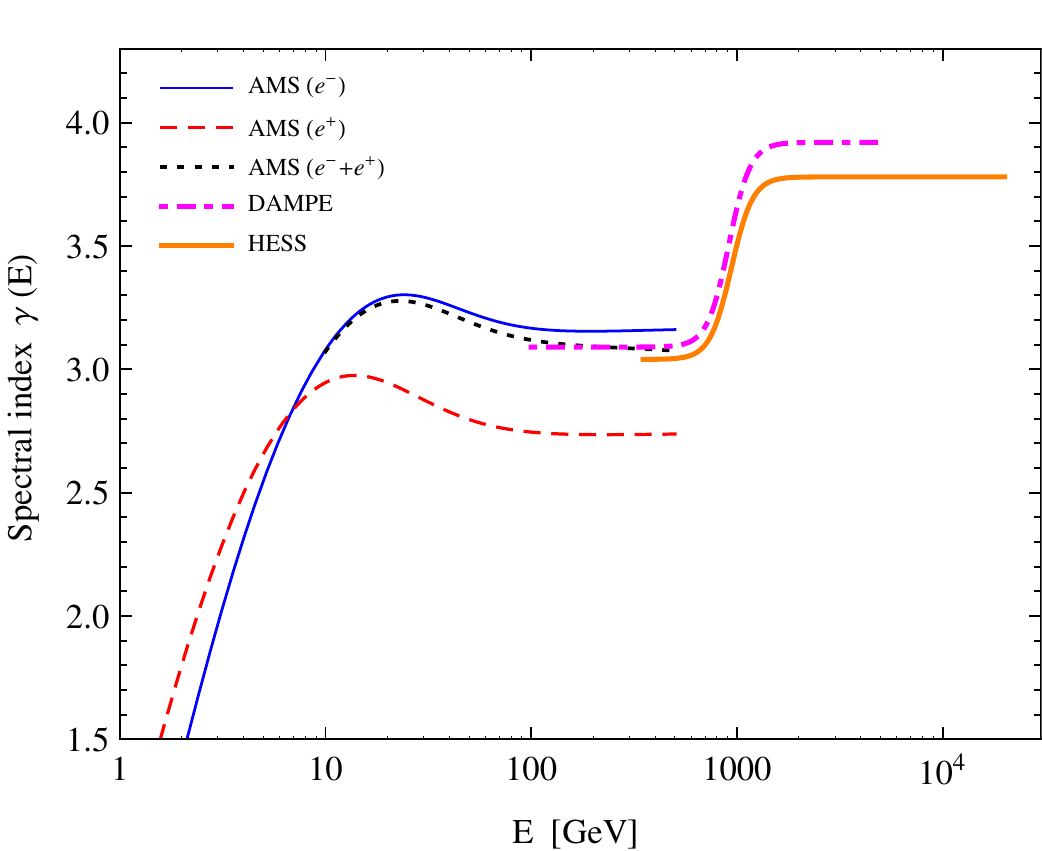}
\end{center}
\caption {\footnotesize
 Spectral indices of the lepton fluxes.
 The lines are calculated using fits to the data.
 The fits to the AMS02 data discussed in the text. The fits to the DAMPE
 and HESS data are from \cite{Ambrosi:2017wek} and \cite{hess-icrc2017}.
\label{fig:index_lepton} }
\end{figure}

A useful quantity to study the properties of a CR spectrum is
the (energy dependent) spectral index, defined in Eq.~(\ref{eq:spectral-index}).
Figure~\ref{fig:index_lepton} shows the spectral indices
of the $e^-$, $e^+$ and $(e^- + e^+)$ fluxes
calculated from fits to the data.
For the AMS02 data we have used the best fits functions presented in this work,
for the DAMPE \cite{Ambrosi:2017wek} and HESS \cite{hess-icrc2017}.
data the fits obtained in the original publications

The main points that emerge from a study of the measurements
of the electron and positron spectra are the following:

\begin{enumerate}

\item For $E \lesssim 20$~GeV the $e^-$ and $e^+$ spectra have a ``curved'' shape,
 with spectral indices that change continuosly. In this energy range the effects
 of solar modulations is important, and the estimate of the interstellar spectra
 model dependent.
 In this work we have shown that if the solar modulations are modeled
 with the FFA approximations, the interstellar spectra for
 both electrons and positrons are consistent with unbroken power laws.
 This is an intriguing result that clearly
 requires a more detailed study.
 
 \item 
 The spectral indices for $e^-$ and $e^+$ have a similar 
 behaviour in the energy range 10--40~GeV, reaching a maximum 
 at $E \simeq 20$~GeV, and then reaching an approximately constant value
 (with a step $\Delta \gamma \simeq 0.15$ for electrons and 0.24 for positrons).
 The existence of this structure was first noted by the AMS02 collaboration
 \cite{ams02-electrons-positrons-2014}.
 Both spectra have therefore gradual hardenings at $E \sim 30$~GeV, generated
 by (a) mechanism(s) that need to be clarified
 (more discussion in  appendix~\ref{sec:hardening}).

\item In the energy range 40--400~GeV both the $e^-$ and $e^+$ spectra
 can be reasonably well described by simple power laws.
 Electrons have a significantly softer spectrum 
 ($\gamma_{e^-} \simeq 3.17$, $\gamma_{e^+} \simeq 2.74$).
 The electron spectrum is much steeper than the proton one.

\item The spectrum of the sum $(e^- + e^+$) exhibit a sharp and
 large break at $E \simeq 1$~TeV.
 This (together with the ``GZK cutoff'')
 is perhaps the most prominent
 spectral structure is all of the cosmic ray data.

\item From the published data, it is not possible to reach a firm conclusion
 about the spectral shape of the separate spectra of $e^-$ and $e^+$
 above $E \gtrsim 500$~GeV.

\item At $E \approx 1$~TeV, the sum $(e^- + e^+$) is dominated by electrons
 (extrapolating the lower energy data as unbroken power laws one estimates a
 ratio $e^-/e^+ \simeq 3$--4).
 Therefore, the measurements of the $(e^- + e^+$) flux imply that the electrons
 have a spectral break at $E \simeq 1$~TeV.
 On the other hand, one can also conclude the also the positron spectrum cannot
 continue as an unbroken power law, and must undergo a softening
 at $E \lesssim 1.5$~TeV, because in the absence of such a break
 the $e^+$ flux would emerge as the dominant component of the ($e^-+e^+$) spectrum in contrast
 to the observation (see Fig.~\ref{fig:allele2}).
 The $e^+$ spectral break could develop at the same energy as for electrons,
 but could also be at lower (perhaps as low as 600~GeV)
 or higher energy.

\item At present, the measurements of DAMPE extend to 4.5~TeV, and those of HESS to 20~TeV,
 and the data above the break at 1~TeV appear to be consistent with a power law spectrum.
 This is in fact a significant constraint for the modeling of the CR sources.
\end{enumerate}

The question we want to investigate is where in the electron and positron
spectra it is possible to identify signatures associated to energy losses.
To address this question we will study some simple models of CR propagation
to see what kind of effects one could expect to see.

\section{Theoretical Models for the Cosmic Ray spectra}
\label{sec:models}

\subsection{Formation of the CR Spectra} 
\label{sec:formation}
In general, the flux of cosmic rays of type $j$ and energy $E$
that are observed at the point $\vec{x}$ at time $t_{\rm obs}$ can be written as the
convolution:
\begin{eqnarray}
 \phi_j (E, \vec{x}, t_{\rm obs})
 & = &
 \frac{ \beta \, c}{4 \, \pi} ~
 n_j (E, \vec{x}, t_{\rm obs}) \nonumber \\[0.22 cm]
 & = &
 \frac{ \beta \, c}{4 \, \pi} ~
 ~\int_{-\infty}^{t_{\rm obs}} ~ dt_i
 ~\int d^3 x_i ~
 ~\int dE_i ~
 q_j (E_i, \vec{x}_i, t_{\rm obs} - t_i) ~
 \mathcal{P}_j (\vec{x}, E, \vec{x}_i, E_i, t_{\rm obs} - t_i) 
\label{eq:flux-general}
\end{eqnarray}
In this equation the factor $\beta \,c/(4 \, \pi)$ relates, assuming isotropy, 
the flux to the particle density $n_j$. The CR density is obtained as the integral
over space, time and initial energy of an integrand that 
is the product of two fundamental quantities:
the source spectrum $q_j (E_i, \vec{x}_i, t_i)$ and the
propagation function $\mathcal{P}_j (\vec{x}, E, \vec{x}_i, E_i, t)$.
The source spectrum (with dimensions $L^{-3} \, T^{-1} \, E^{-1}$)
is the rate of particles of type $j$ and energy $E_i$ that are released
in interstellar space per unit volume around the point $\vec{x}_i$ at the time $t_i$.
The propagation function (with dimensions $L^{-3} \, E^{-1}$)
expresses the probability that a particle of type $j$,
created with energy $E_i$ at the point $\vec{x}_i$ and t the time $t_i$, is
then found at the time $t_i + t$ with energy $E$ at the point $\vec{x}$.

Equation~(\ref{eq:flux-general}) is a general expression
for the CR fluxes, valid for most types of models.
The most important non trivial hypothesis that enters into the equation
is the assumption that (with respect to CR propagation),
the Galaxy is in a stationary state,
so that the propagation function
$\mathcal {P}_j$ is only a function of the time difference $t = t_{\rm obs} - t_i$.

In most cases Eq.~(\ref{eq:flux-general}) is only a formal solution
for the formation of the CR fluxes, because it can be very difficult
to compute the propagation functions, but in the following
we will discuss two simple models where the
functions $\mathcal{P}_j$ have exact analytic expressions.

Integrating the propagation function over the $\vec{x}$ and $E$
one obtains the probability that a particle
(created at the point $\vec{x}_i$ with energy $E_i$) is still confined in the Galaxy
after a time $t$:
\begin{equation}
\int d^3 x ~\int dE ~
 \mathcal{P}_j (\vec{x}, E, \vec{x}_i, E_i, t)= P_{\rm surv} (E_i, \vec{x}_i, t) ~.
\end{equation}
This survival probability must be in the interval [0,1],
and decreases monotonically with $t$,
with the (physically obvious) limits:
$P_{\rm surv} = 1$ for $t \to 0$, and $P_{\rm surv} = 0$ for $t \to \infty$. 

The average residence time in the Galaxy for a particle created at the point
$\vec{x}_i$ with energy $E$ is then:
\begin{equation}
  \langle t_{\rm res} (E, \vec{x}_i) \rangle = \int_0^\infty dt~ t
  ~\left | \frac{dP_{\rm surv} (E, \vec{x}_i, t)}{dt}
 \right |~.
\label{eq:t-residence}
\end{equation}

\subsection{Energy losses}
\label{sec:losses}
The main mechanisms of energy losses for relativistic electrons and positrons
traveling in interstellar space are synchrotron radiation and Compton scattering,
and the rate of energy loss is reasonably well approximated with the form:
\begin{equation}
- \frac{dE}{dt} \equiv \beta(E) \simeq b \, E^2 \simeq 
 \frac{4}{3} \; \sigma_{\rm Th} \, c \;
 \left [\rho_B + \; \rho_\gamma^* (E) \right ]\; \frac{E^2}{m_e^2} ~.
\label{eq:eloss}
\end{equation}
In this expression $\sigma_{\rm Th}$ is the Thomson cross section,
$\rho_B = B^2/(8 \pi)$ is the energy density stored in the magnetic field,
and $\rho_\gamma^*(E)$ is the energy density in target photons
with energy $\varepsilon \lesssim m_e^2/E$.
This kinematical constraint insures that the $e \gamma$ scatterings
are in the Thomson regime, excluding
collisions at higher center of mass energy (in the Klein--Nishina regime)
where the cross section is suppressed.

Other mechanisms, such as bremsstrahlung and ionization, give smaller contribution
to the energy loss, 
and it is a reasonable approximation to neglect them, and use the 
simple expression of Eq.~(\ref{eq:eloss}) that grows quadratically with energy.

Because of the dependence on $\rho_B$ and $\rho_\gamma$, the rate of energy loss
is not constant in space. In this work however we will make the
approximation to consider the Galactic confinement volume as homogeneous,
taking for $\rho_B$ and $\rho_\gamma$ a constant average value.

In the vicinity of the solar system, the magnetic field
is of order $B \simeq 3$~$\mu$G, with a corresponding energy density
$\rho_B \simeq 0.22$~eV/cm$^{-3}$. The magnetic field is larger
near the Galactic center, but also falls rapidly 
with the distance from the Galactic equatorial plane.
The interstellar radiation field is formed by three components:
the cosmic microwave background radiation (CMBR), stellar light and
dust emitted infrared radiation.
The CMBR fills homogeneously all space
with an energy density $\rho_{\rm CMBR} \simeq 0.260$~eV/cm$^3$
formed by very soft photons (with average energy $\langle \varepsilon \rangle \simeq 6.3 \cdot 10^{-4}$~eV),
that are an effective target also for high energy photons.
Stellar light and infrared radiation have, in the solar neighborhood,
an energy density of order 0.5~eV/cm$^3$, that however changes rapidly with in space.
In the following, when needed, we will use as a first order estimate
the value $\langle \rho_B + \rho^*_\gamma \rangle \simeq 0.5$~eV/cm$^{3}$.
One should keep in mind that this average
depends on the size and shape of the CR confinement volume that are very poorly known.
It is straightforward to rescale the results for a different
estimate if so is desired.

From the rate of energy loss it is possible to compute a
characteristic time for energy loss:
\begin{equation}
 T_{\rm loss} (E) = \frac{E}{|dE/dt|} \simeq \frac{1}{b \, E}
\label{eq:tloss0}
\end{equation}
In the last equality we have used the quadratic form
for the energy loss $\beta (E) = b \, E^2$, this results in
the dependence $T_{\rm loss} \propto E^{-1}$.
A numerical estimate is:
\begin{equation}
 T_{\rm loss} (E) \simeq
621.6 ~
\left [ \frac{0.5~{\rm eV}\;{\rm cm}^{-3}}
 {\left \langle \rho_B + \rho_\gamma^* \right \rangle}
\right ]
\left [ \frac{\rm GeV}{E} \right ]~
 ~{\rm Myr}~.
\label{eq:loss-num}
\end{equation}

If the energy loss is considered as a stationary, continuous process,
that does not depend from the space coordinates,
the energy of a particle is a well determined function of time.
The three quantities \{$E_i$, $E_f$, $t$\} (that is the
initial and final energy of a particle that propagates
for a time $t$) are related by the equation:
\begin{equation}
 \int_{E_f}^{E_i} ~\frac{dE^\prime}{\beta(E^\prime)} = t
 \label{eq:eloss1}
\end{equation}
This equation can be solved to obtain any one of the three quantities
as a function of the other two.
For $\beta (E) = b \, E^2$ one has simple explicit expressions:
\begin{equation}
E_f(E_i, t) = \frac{E_i}{1 + b \; E_i \;t} ~,
\label{eq:ef}
\end{equation}
\begin{equation}
E_i(E_f, t) = \frac{E_f}{1 - b \; E_f \, t} 
\label{eq:ei}
\end{equation}
[for the assumptions made one must have $E_f(E,t) = E_i (E, -t)$], and 
\begin{equation}
t(E_i,E_f) = \frac{1}{b \, E_f} - \frac{1}{b \, E_i}
\label{eq:te}
\end{equation}
The limit of $t(E_i, E_f)$ for $E_i \to \infty$ is finite,
indicating that electrons and positrons
observed at the Earth with energy $E_f$ have a maximum age:
\begin{equation}
 t_{\rm max} (E_f) =
\lim_{E_i \to \infty} t(E_i, E_f) = 
 \int_{E_f}^\infty \frac{dE^\prime}{\beta(E^\prime)} \simeq
 \frac{1}{b \, E_f} ~.
\label{eq:tmax}
\end{equation}

\subsection{The Leaky Box model}
\label{sec:leaky} 
The simplest model to describe the Galactic cosmic rays spectra
is the so called Leaky Box model.
In the model the space dependence of the cosmic ray fluxes and sources
are neglected, and each CR species is entirely described
by the function $n(E,t)$ that gives the density of the particles
as a function of energy and time
(in the following the subscript $j$ that indicates the particle type will be left implicit).
The time evolution of $n(E,t)$ is controled by the equation:
\begin{equation}
 \frac{\partial n(E,t)}{\partial t} =
 q(E,t) - \frac{n(E,t)}{T_{\rm esc} (E)} + \frac{\partial}{\partial E}
 \left [ \beta (E) \; n(E, t) \right ] ~.
\label{eq:leaky-definition}
\end{equation}
The spectrum of one particle type
is therefore entirely determined by three functions:
the source spectrum $q(E,t)$, the escape time $T_{\rm esc} (E)$ and the rate of energy loss $\beta(E)$.

The propagation function (that obviously cannot depend on the space coordinates),
is determined by $T_{\rm esc} (E)$ and $\beta(E)$, and has the form:
\begin{equation}
 \mathcal{P} (E, E_i, t) = P_{\rm surv} (E_i, t) ~\delta [E - E_f (E_i, t)] ~.
\label{eq:leaky-general}
\end{equation}
In this expression the delta function expresses the fact that
(using the assumption of continuous energy loss)
a CR particle has a well defined energy at all times.
The function $P_{\rm surv} (E_i, t)$ is the probability
that a CR particles released in interstellar space with energy $E_i$ is still
in the Galaxy after a time interval $t$, and can be calculated as:
\begin{equation}
 P_{\rm surv} (E_i, t) =
 \exp \left [
- \int_0^t \frac{dt^\prime}{T_{\rm esc} [E_f (E_i,t^\prime)]} 
\right ]
 \end{equation}
If energy losses are negligible (that is in the limit $\beta(E) =0$),
the survival probability becomes a simple exponential $P_{\rm surv} (E,t) = e^{-t/T_{\rm esc} (E)}$.
In this case $T_{\rm esc}(E)$ is also exactly equal
[see Eq.~(\ref{eq:t-residence})] to the average residence time of the particles.

The CR density can be calculated using the general expression of Eq.~(\ref{eq:flux-general}).
The integration over $E_i$ can be performed
using the delta function, obtaining a general solution
in the form of one single integral over time:
\begin{equation}
 n(E) 
 = \int_0^{t_{\rm max}(E)} dt~ \frac{\beta[E_i (E,t)]}{\beta(E)} ~
 q[E_i(E,t), t] ~P_{\rm surv} [E_i (E,t), t] ~,
\label{eq:leaky-t1}
\end{equation}
where $t_{\rm max} (E)$ is the maximum residence time in the Galaxy for a particle
 observed with energy $E$ [see Eq.~(\ref{eq:tmax})], and $E_i (E, t)$ is the past energy of a particle
observed with energy $E$ [see Eq.~(\ref{eq:ei})].

If the source spectrum is constant in time and energy lossese are negligible
(so that $E=E_i$ with the survival probability is a simple exponential):
the time integration is trivial, and one obtains the simple and obvious result 
\begin{equation}
 n(E) = q(E)~T_{\rm esc} (E) ~.
\label{eq:noloss0}
\end{equation}

If the CR source is stationary, and the escape probability is negligible,
(that is in the limit $T_{\rm esc} \to \infty$),
the survival probability is unity, and the CR spectrum becomes:
\begin{equation}
n(E) = \frac{1}{\beta(E)} ~\int_E^{\infty} dE_i ~q(E_i)
\label{eq:asymptotic}
\end{equation}

A phenomenologically interesting case is when:
\begin{itemize}
\item The source spectrum is stationary in time,
 confined to the Galactic plane, constant and with a simple power law form:
 \begin{equation}
 q(E, \vec{x}) = q_0 \; E^{-\alpha} ~\delta [z]
\label{eq:q-diff}
 \end{equation}
\item The escape time is also a power law: $T_{\rm esc} (E) = T_0 \; E^{-\delta}$.
\item The rate of energy loss is quadratic in energy ($\beta(E) = b\, E^2$),
 and therefore the loss time is $T_{\rm loss} = 1/(b E)$.
\end{itemize}
In this situation the critical energy $E^*$
defined by the condition that the escape and loss time are equal,is given by:
\begin{equation}
E^* = (T_0 \, b)^{1/(\delta -1)} ~.
\label{eq:estar}
\end{equation}
The form of the CR spectrum in the limits of low and high energy
can be obtained from the more general expressions of 
Eqs.~(\ref{eq:noloss0}) and~(\ref{eq:asymptotic}).
These asymptotic solutions are power laws 
with exponents $(\alpha + \delta)$ at low energy and $\alpha +1$
at high energy.
\begin{equation}
 n(E) = \begin{cases}
 {q_0 \, T_0} ~E^{-(\alpha + \delta) }
 & \text {for $E \ll E^*$} ~, \\[0.30 cm]
 {q_0}/[{b \, (\alpha-1)}] ~E^{-(\alpha + 1) }
 & \text {for $E \gg E^*$} ~.
 \end{cases}
\label{eq:step0}
\end{equation}
The general solution can be written in the form:
\begin{equation}
 n(E) = (q_0 \, T_0) ~E^{-(\alpha + \delta)} ~ F_{\rm loss}^{\rm LB} (a)
\end{equation}
that is as the product of the no energy loss solution
multiplied by a correction factor that can be expressed as
a function of the energy dependent parameter $a$
\begin{equation}
 a = \frac{T_{\rm esc} (E)}{T_{\rm loss} (E)}
= \left ( \frac{E}{E^*} \right )^{1-\delta} ~.
\end{equation}
The function $F_{\rm loss}^{\rm LB} (E)$ is given by:
\begin{equation}
F_{\rm loss}^{\rm LB} (a) =
 \int_0^{1/a} ~d\tau ~(1- a \, \tau)^{\alpha -2} ~
 \exp \left [ -\frac{1}{a \; (1- \delta)} \; [1 - (1 - a \, \tau)^{1-\delta}]
 \right ] ~.
\label{eq:knee-LB}
\end{equation}
It is straightforward to check that in the limit $a \to 0$ one has
$F_{\rm loss}^{\rm LB} \to 1$; while for $a \gg 1$ one has
\begin{equation}
F_{\rm loss}^{\rm LB} (a) \simeq [a \, (\alpha-1)]^{-1} = (\alpha-1)^{-1} \; (E/E^*)^{\delta-1} ~,
\end{equation}
recovering the asymptotic expressions for
$E \ll E^*$ and $E \gg E^*$ in Eq.~(\ref{eq:step0}).

\begin{figure}[bt]
\begin{center}
\includegraphics[width=11.0cm]{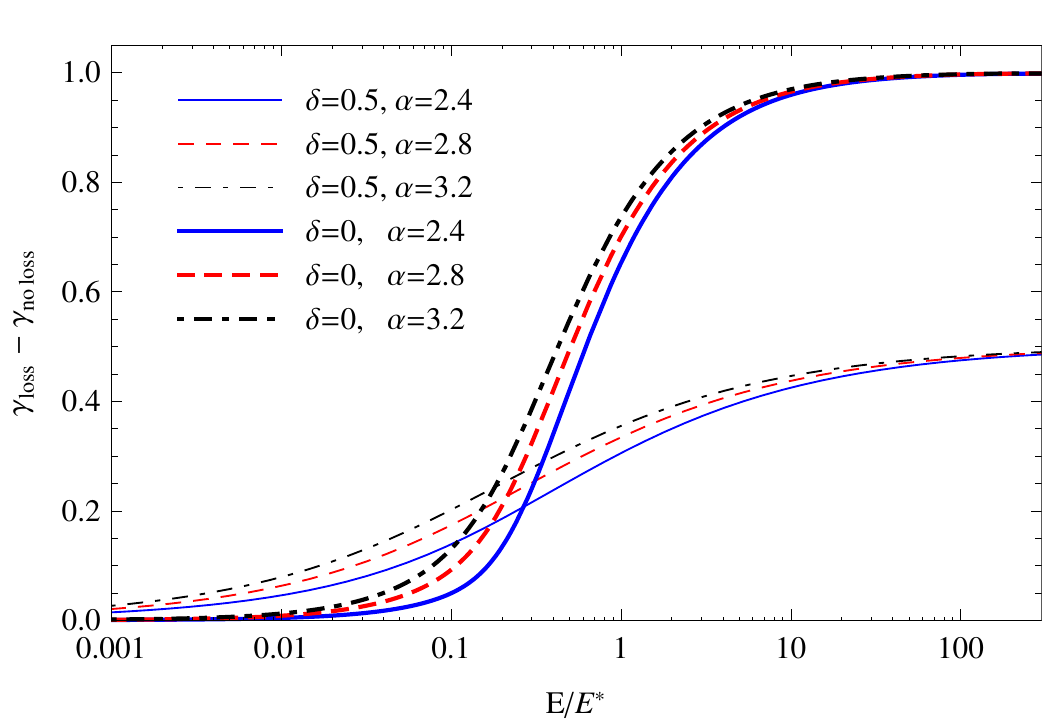}
\end{center}
\caption {\footnotesize
Spectral feature generated by energy losses
calculated in a Leaky Box model. The source
spectrum and the escape time have the energy dependence $q(E) \propto E^{-\alpha}$
and $T_{\rm esc} (E) \propto E^{-\delta}$.
\label{fig:leaky} }
\end{figure}

Having found an exact expression for the CR spectrum
it is straightforward to compute the energy
dependence of the spectral index $\gamma(E)$.
The effect of energy losses is to
generate a ``step'' in the index of size ($\Delta \gamma = 1-\delta)$,
as the spectrum slope changes from the value $(\alpha + \delta)$ at low
energy to the values $\gamma = (\alpha+1)$ at high energy.
Some examples of the spectral feature generated by energy losses, calculated 
for the set of assumptions outlined above, are given in Fig.~\ref{fig:leaky}.

The step in the spectral index is centered at an energy proportional to
$E^*$ [defined in Eq.~(\ref{eq:estar})], but approximately a factor of two smaller.
The width of the energy range where the step develops is very broad
and depends strongly on $\delta$ (and more weakly also on $\alpha$).
For $\delta \simeq 0.5$ the step $\Delta \gamma$ 
develops when the particle energy grows by a factor of
approximately 30, and this corresponds to a width $w$ is of order 1.6 to 1.7
(depending on $\alpha$).
For $\delta \simeq 0$, the step in spectral index develops
when the energy grows by a factor of order 5,
and the width $w$ is of order 0.4 to 0.6.

\subsection{Diffusion model}
\label{sec:diffusion}
The idea to describe the propagation of cosmic rays in the Milky Way as diffusion
was introduced by Morrison, Olbert and Rossi in the 1950's \cite{rossi-1954},
then discussed extensively by Ginzburg and Syrovatskii \cite{Ginzburg-1964}
and is at present the basis for essentially all models for the propagation of CR in the Galaxy. 

In this work we will use the simplest possible diffusive propagation model, 
where the CR confinement volume is taken the
(infinitely large) region of space with $|z| \le H$
(with the $z$ plane coincident with the Galactic equatorial plane).
In the confinement volume the effects
of the magnetic field are described as isotropic diffusion with
a diffusion coefficient $D(E)$ that is a function of rigidity,
but is independent from position\footnote{In this work we only consider ultrarelativistic particles
 of unit electric charge, so that $p/|Z| \simeq E$, and for simplicity,
 we will use the notation $D(E)$ for the diffusion coefficient
 (that depends of the absolute value of the rigidity and on the particle velocity $\beta$).}.
The two surfaces $z=\pm H$ are considered as absorption
barriers, and particles that reach these surfaces leave permanently the Galaxy.
Particle (of a given type) in the diffusive volume lose energy
with a space independent rate described by $\beta(E)$.
The model is therefore entirely defined by the
``halo size'' $H$ and by the two functions $\beta(E)$ and $D(E)$.

Some important quantities in the model are the loss time $T_{\rm loss} (E)$,
the diffusion or escape time $T_{\rm diff} (E)$,
the critical energy $E^*$,
the diffusion radius $R (E_i, E_f,t)$ and the
maximum diffusion radius $R_{\rm loss} (E)$.

The loss time has been defined in Eq.~(\ref{eq:tloss0}),
the diffusion time:
\begin{equation}
 T_{\rm diff} (E) \equiv T_{\rm esc} (E) = \frac{H^2}{2 \, D(E)} ~,
\label{eq:tdiff} 
\end{equation}
is the time after which a particle of constant energy $E$,
diffusing in a homogeneous medium with diffusion coefficient $D(E)$
travels an average distance
$\langle x^2 \rangle = \langle y^2 \rangle = \langle z^2 \rangle = H^2$.
It can be demonstrated \cite{Lipari:2014zna} that this quantity is equal to 
the average escape time from the Galaxy for a particle of energy $E$ created
on the Galactic plane $z =0$.
The critical energy $E^*$ is determined by the condition
$T_{\rm loss} (E^*) = T_{\rm esc} (E^*)$.

The quantity $R^2 (E_i, E_f, t)$ is equal to the
average square distance 
($R^2 = \langle x^2 \rangle = \langle y^2 \rangle = \langle z^2 \rangle$)
traveled by a particle of initial energy $E_i$ (and final energy $E_f$)
propagating in a homogeneous medium with diffusion coefficient $D(E)$ for a time $t$.
The propagation includes the effects of energy losses
[determined by the function $\beta (E)$], but neglects absorption effects.
The quantity can be calculated performing one the three integrals:
\begin{equation}
 R^2 (E_i, E_f, t)
 = 2 \; \int_0^t dt^\prime ~D[E_i (E_f,t^\prime)]
 = 2 \; \int_0^t dt^\prime ~D[E_f (E_i,t^\prime)]
 = 2 \; \int_{E_f}^{E_i} ~dE^\prime ~ \frac{D(E^\prime)} {\beta (E^\prime)} ~.
\label{eq:R2-definition}
\end{equation}
The three quantities $E_i$ and $E_f$ and $t$ are not independent,
because they are related by Eq.~(\ref{eq:eloss1}),
therefore $R$ can be expressed as function of any pair
of variables out of the set $\{E_i,E_f, t\}$.
In the limit of negligible energy losses one has $E_i = E_f$,
and $R^2(E,t) = 2 \, D(E) \, t$.

The maximum diffusion radius $R_{\rm loss} (E)$ is the maximum value of $R$ for
particles with final (observed) energy $E$, and corresponds to the limit
for $E_i \to \infty$ of $R(E_i, E, t)$.
This quantity is finite if the rate of energy loss grows sufficiently
rapidly with $E$.

In the following we will consider again the phenomenologically interesting case where
the rate of energy loss is quadratic in energy ($\beta(E) = b \, E^2$) 
and the diffusion coefficient is a power law
with exponent $\delta$ ($D(E) = D_0 \; E^{\delta}$).
In this situation the loss time has the form $T_{\rm loss} (E) = 1/(b \, E)$,
and the diffusion time is also a power law ($T_{\rm esc} (E) = (H^2/(2 \, D_0)) ~E^{-\delta}$).
This implies that the critical energy $E^*$ is:
\begin{equation}
E^* = \left ( \frac { H^2 \, b}{2 \, D_0} \right )^{1/(\delta-1)} ~.
\label{eq:estar-diff}
\end{equation}
The maximum diffusion radius is finite and has the form:
\begin{equation}
 R_{\rm loss}^2 (E) = \frac{2 \, D_0}{b \, (1-\delta)}
 ~E^{-(1-\delta)} =
 \frac{H^2}{1 -\delta} ~\frac{T_{\rm loss} (E)}{T_{\rm esc} (E)} 
 = \frac{H^2}{1 -\delta} ~
 \left ( \frac{E}{E^*} \right )^{\delta -1} ~.
\label{eq:r2maxa}
\end{equation}
If (as it is expected) $\delta < 1$
the maximum diffusion radius decreases with energy $\propto E^{-(1-\delta)/2}$.

For a finite value of $E_i$ [or for a time $t < t_{\rm max}(E)$], the diffusion radius is:
\begin{equation}
 R^2 (E_i, E_f, t)
 = R_{\rm loss}^2 (E_f) ~ 
 \left [1 - \left (1- \frac{t}{t_{\rm max} (E)} \right )^{1-\delta} \right ]
 = R_{\rm loss}^2 (E_f) ~ 
 \left [1 - \left ( \frac{E_f}{E_i} \right )^{1-\delta} \right ]
\label{eq:rdiff1} 
\end{equation}

In the simple diffusive model that we have just described, 
the propagation function can be written explicitely:
\begin{equation}
 \mathcal{P} (\vec{x}, E, \vec{x}_i, E_i, t) =
\frac{1}{2 \, \pi \, R^2} ~e^{-r^2/(2 \, R^2)} ~\frac{1}{H} \;
g \left ( \frac{z}{H}, \frac{z_i}{H}, \frac{R^2}{H^2} \right) 
~\delta [E - E_f (E_i, t)] ~
\label{eq:prop-diffusion}
\end{equation}
where $R$ is the diffusion distance defined above in Eq.~(\ref{eq:R2-definition}).

Inspecting Eq.~(\ref{eq:prop-diffusion}) one can see that the propagation along the
$x$ and $y$ directions is described by a simple gaussian,
while in the $z$ direction one has to take into account
the presence of the absorption planes at $z = \pm H$, and
the propagation has a more complicated expression, encoded in the function $g$.

The solution of the diffusion equation in the presence
of the two absorption barriers can be obtained
(see for example \cite{cox-miller})
as the solution without the barriers but including together with
the real source at point $x_i$ an infinity of ``mirror''
sources and sinks located symmetrically in the unphysical
region outside the region between the two barriers.

\begin{figure}[bt]
\begin{center}
\includegraphics[width=11.0cm]{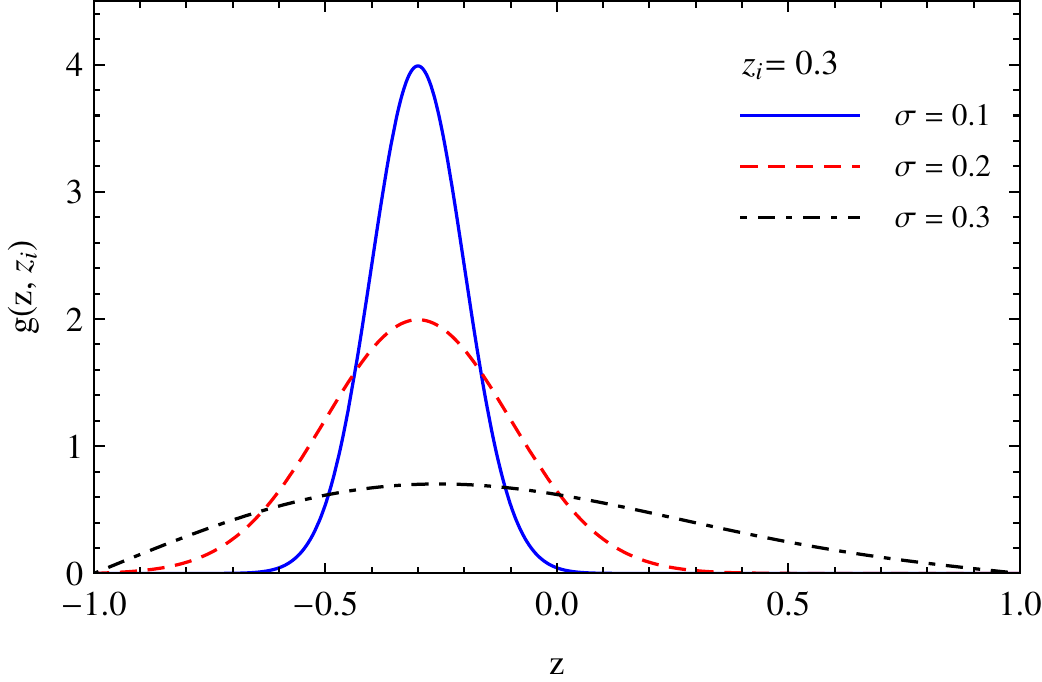}
\end{center}
\caption {\footnotesize
Examples of the function $g(x,x_i,\sigma^2)$.
\label{fig:g_example} }
\end{figure}

Putting the absorption barriers at $\pm 1$, and
using adimensional variables for $x$, $x_i$ and $\sigma^2$,
the solution takes the form:
\begin{equation}
 g(x,x_i, \sigma^2) =
 \frac{1}{\sqrt{2 \, \pi} \; \sigma}
~ \sum_{n=-\infty}^{+\infty} 
\left [
 e^{- [x - x_n^+ (x_i) ]^2/ (2 \, \sigma^2)} -
 e^{- [x - x_n^- (x_i) ]^2/ (2 \, \sigma^2)}
 \right ] ~.
\label{eq:gdef}
\end{equation}
This expression contains infinite sources located at the points:
\begin{equation}
x_n^+ = x_i + 4 \, n ~,
\label{eq:n-sources}
\end{equation}
(with $x_0^+ = x_i$ the real source), 
and infinite sinks at the points:
\begin{equation}
x_n^- = -x_i + 4 \, n + 2 ~.
\label{eq:n-sinks}
\end{equation}
The solution for the problem where the absorption barriers at $z= \pm H$ can be obtained
simply rescaling the expression in Eq.~(\ref{eq:gdef}). 
Some numerical examples of the function $g(x,x_i, \sigma^2)$ are
shown in Fig.~\ref{fig:g_example}.
The function is always positive in the interval $[-1,1]$,
vanishing at the boundary points ($x = \pm 1$). 
For $\sigma^2 \ll 1$ the function is in good approximation
a simple gaussian of unit normalization and width $\sigma$
centered at the point $x = x_i$. 

A useful property of the function $g$ is that:
\begin{equation}
 \int_0^\infty d\tau ~g(x,x_i, \tau) =
 \begin{cases}
(1+x) \; (1-x_i) ~~~& \text{for } x < x_i \\[0.12 cm]
(1-x) \; (1+x_i) ~~~& \text{for } x > x_i 
 \end{cases}
\label{eq:g-int}
\end{equation}

Having an explicit analytic expression for the propagation function,
it is straightforward to compute the CR spectrum
for arbitrary forms of the energy loss function $\beta(E)$
and the diffusion coefficient $D(E)$
and for an any source spectrum $q(E, \vec{x}, t)$
performing numerically the integrations in Eq.~(\ref{eq:flux-general}).

If one makes the assumptions that: 
\begin{itemize}
\item [(i)] The observation point is on the plane $z=0$.
\item [(ii)] The source spectrum is constant in time, confined
 to the plane $z=0$ and homogeneous in the plane,
\end{itemize}
then the integration over space and energy in Eq.~(\ref{eq:flux-general})
are trivial, and the CR density can be expressed as an integral over time:
\begin{equation}
 n(E) = \frac{1}{H} ~\int_0^{t_{\rm max} (E)} dt
 ~\frac{\beta [E_i (E,t)]}{\beta (E)} 
 ~q\left [E_i (E, t), t \right ] ~
 g\left ( 0,0, \frac{R^2[E_i (E,t), t)} {H^2} \right ) ~.
\label{eq:sol-diffusive}
\end{equation}
The last integration must in general be performed numerically,
however in the limit where the energy loss is negligible,
that is for $\beta (E) = 0$ when 
$E_i = E$, $t_{\rm max} (E) \to \infty$ and $R^2 (E,t) = 2 \, D(E) \, t$,
making use of Eq.~(\ref{eq:g-int}) the integration can be done analytically 
with the result $n(E) = q(E) \, T_{\rm esc}(E)/H$.

If one makes the additional assumptions that:
\begin{itemize}
\item [(iii)] The source spectrum is a power law: $q(E) = q_0 \; E^{-\alpha}$.
\item [(iv)] The diffusion coefficient is also a power law: $D(E) = D_0 \, E^{\delta}$
\item [(v)] The rate of energy loss is quadratic in energy: $\beta(E) = b \, E^2$. 
\end{itemize}
the CR spectrum in the limit of no energy loss takes the power law form:
\begin{equation}
 n(E) \simeq n_0 (E)
 = \frac{q(E) \,T_{\rm esc} (E)} {H} 
 = \frac{q_0 \; H}{2 \, D_0} ~E^{-(\alpha + \delta)} 
\label{eq:n-0}
\end{equation}
An exact solution can also be found if 
escape of the Galaxy can be neglected
(this corresponds to the limit $H\to \infty$, or $T_{\rm loss}/T_{\rm esc} \ll 1$).
In this limit the function $g$, that describes propagation in the $z$ direction,
becomes simply a gaussian, and the CR density takes the form
\begin{equation}
 n(E) \simeq n_{\rm loss} (E) =
k(\alpha, \delta)~
 \frac{q(E) ~T_{\rm loss} (E)}{R_{\rm loss} (E)} \;
 = k(\alpha,\delta) \, \sqrt{1-\delta} ~\frac{q_0}{\sqrt{2 \, D_0 \, b}} 
 ~E^{-(\alpha + (1+\delta)/2)} 
 \label{eq:n-loss}
\end{equation}
where $k(\alpha,\delta)$ an adimensional constant
of order unity that depends on the exponents $\alpha$ and $\delta$:
\begin{equation}
k(\alpha, \delta)
 = \frac{1}{\sqrt{2 \, \pi}} ~\int_0^1 d\tau^\prime ~
 \frac{ (1-\tau^\prime)^{\alpha-2}}{\sqrt{1 - (1-\tau^\prime)^{1-\delta}}}
\label{eq:kalpha}
\end{equation}
In the general case, the CR spectrum can be written in the form:
\begin{equation}
 n(E) = n_0 (E) ~F_{\rm loss}^{\rm diff} (a) = \frac{q(E) ~T_{\rm esc} (E)}{H} ~~F_{\rm loss}^{\rm diff} (a) 
\end{equation}
as the product of the no--energy loss expression times a correction factor
that depends on the ratio $E/E^*$ via the parameter $a$:
\begin{equation}
a = \left ( \frac{E}{E^*} \right )^{1-\delta} = \frac{T_{\rm diff(E)}}{T_{\rm loss} (E)} ~.
\end{equation}
The function $F_{\rm loss}^{\rm diff} (a)$ has the form:
\begin{equation}
F_{\rm loss}^{\rm diff} (a) = 
 \int_0^{1/a}
 d\tau ~ (1- a \, \tau)^{\alpha-2} \, g[0,0, \sigma^2(a, \tau)]
\label{eq:feature-diff}
\end{equation}
with:
\begin{equation}
 \sigma^2 (\tau, a) =
 \frac{1}{(1-\delta) \, a} ~[ 1- (1 - a \, \tau)^{1-\delta}]
\end{equation}
It is straightforward to see that in the limit $a \to 0$
one has $F_{\rm loss}^{\rm diff} (0) = 1$, while in the limit $a\to \infty$
one finds: $F_{\rm loss}^{\rm diff} \propto a^{-1/2} \propto E^{-(1-\delta)/2}$.
The exact expression is:
\begin{equation}
 \lim_{a \to \infty}
F_{\rm loss}^{\rm diff} (a)
 = k(\alpha, \delta) ~\sqrt{\frac{1-\delta}{a}}
\end{equation}
These results are consistent with the asymptotic behaviors
given in Eqs.~(\ref{eq:n-0}) and~(\ref{eq:n-loss}).
The CR density has spectral index $\alpha + \delta$ at low energy ($E/E^* \ll 1$)
and index $\alpha + (1+\delta)/2$ at high energy ($E/E^* \gg 1$).

\begin{figure}[bt]
\begin{center}
\includegraphics[width=11.0cm]{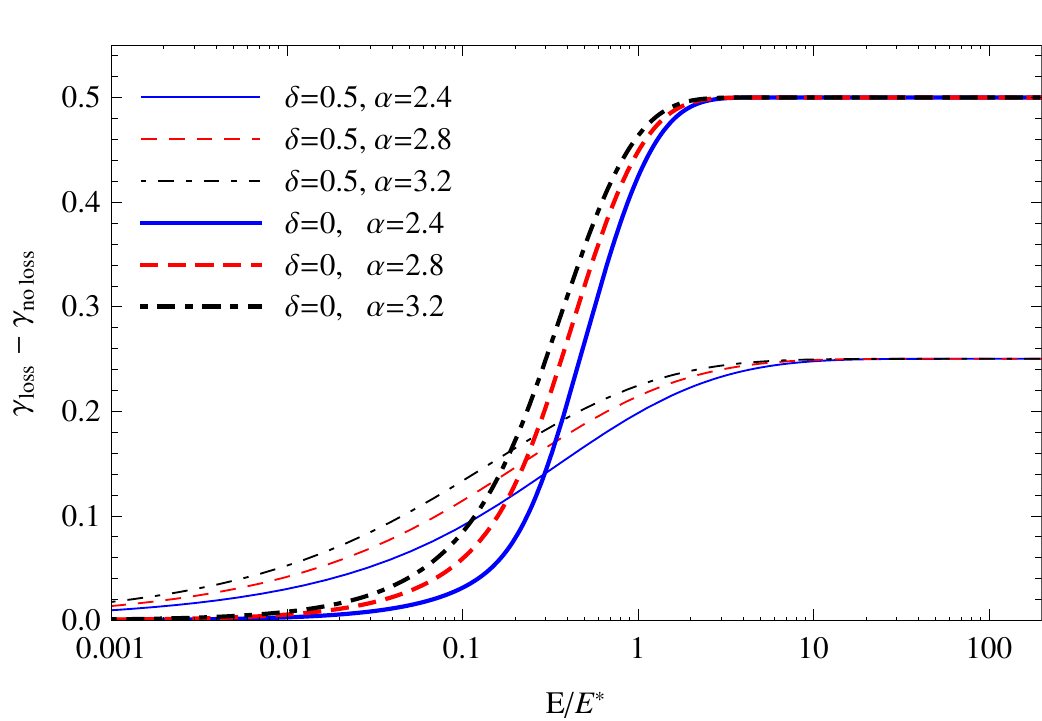}
\end{center}
\caption {\footnotesize
Spectral feature generated by energy losses
in a diffusive model.
The diffusion coefficient is homogeneous in
the finite layer $|z| < H$ and has the energy dependence $D(E) = D_0 \; E^{\delta}$.
The CR source is confined in a thin layer at $z \simeq 0$ and has the form
$q(E) = q_0 ; E^{-\alpha}$.
\label{fig:thin-diffusion} }
\end{figure}

The transition from the low to the high energy behaviour has a shape
that is completely determined by the exponents $\alpha$ and $\delta$.
Some examples of this transition are shown in Fig.~\ref{fig:thin-diffusion}.

The step in the spectral index is again centered at an energy that is proportional
to $E^*$ (but a factor 0.4--0.5 smaller), and develops in a region of
$\log E$ of order 1-2 (a factor 3 to 10). This approximately a factor of two smaller
than in the case of the Leaky Box, but still a rather gradual break.

It is well known that a diffusion model allows not only to compute the flux
as a function of position, but also the dipole $\vec{d}$ of the angular distribution
at each point:
\begin{equation}
 \phi(E, \Omega, \vec{x} ) = \phi (E, \vec{x}) ~
 \left [1 + \vec{d}(E, \vec{x}) \cdot \hat{u}(\Omega) \right ]
\end{equation}
(where $\hat{u}(\Omega)$ is a versor in the direction $\Omega$).
The dipole momentum is antiparallel to the gradient of the CR density
with the value:
\begin{equation}
 \vec{d} (E, \vec{x}) = -\frac{3 \; D(E) }{\beta \, c}
 ~\frac{\vec{\nabla}n(E, \vec{x})}{n(E, \vec{x})}~.
\label{eq:dipole}
\end{equation}
For the simple model we have constructed (with the Solar System exactly on the
Galactic plane, and the CR homogeneously distributed on the plane), the dipole
at the position of the Sun is zero.

\subsection{The energy loss softening feature}
In the discussion above we have   calculated in some  detail the
shape of the CR spectra for electrons and positrons, assuming a very simple
model for the source  (that is an unbroken power law) and  two simple
models  of propagation The main  result is that the observable spectra,
as it is  in fact easy to  predict,  exhibit  a  clear  ``softening feature''
with the spectral index  that   has a ``step'' around the energy
$E^*$.    The calculations  performed above
give  a detailed  description of the
shape of the softening  feature [see Eqs.~(\ref{eq:knee-LB}) and~(\ref{eq:feature-diff})].
Given the simplicity of these models it is however   possible
(in fact likely) that these detailed  predictions  are only an imperfect
description of the real fluxes.
On the other hand, the prediction of the existence
of  a softening feature  (in both $e^-$ and $e^+$  spectra)
should be  considered as a robust prediction, and the identification
of these structures as a very important task.
The shape of  the softening  features (after their identification)
could  then be used  to constrain
the modeling of  CR  propagation in the Galaxy.


\section{Discrete sources}
\label{sec:discrete}
The source of primary Galactic cosmic rays (such as protons and electrons)
can be very likely modeled as the emission from an ensemble
of discrete astrophysical ``events'' that can be approximated as
point--like and short lived for Galactic space and time scales.
Accordingly the flux can be written as the sum of components, in the form:
\begin{equation}
\phi(E) = \sum_k ~\phi_{s, k} (E) ~,
\end{equation}
In the ``standard paradigm'', the source of CR protons and electrons
are supernovae explosions, but other
possibilities (for example Gamma Ray Bursts) have been considered.
If positrons are generated in interstellar space
by standard mechanism of secondary production,
the source is smoothly distributed in space and time, however, 
the possibility that the main source of $e^+$ is acceleration
in astrophysical objects (such as young Pulsars)
has been the object of many studies.

Several authors (see for example \cite{Grasso:2009ma}) have discussed
that if CR of a certain type ($p$, $e^-$, $e^+$, $\ldots$)
are indeed generated in discrete astrophysical events,
the ``granularity'' of the source could have observable consequences.
In this section we discuss this problem,
and show that the discrete nature of the sources
should become manifest for electrons (and also for positrons, 
is they are generated in discrete accelerators)
above a minimum energy (that we will denote $E^\dagger$)
that could be as low as few hundred GeV 
(depending on the properties of the CR sources and propagation).
The effects of the source granularity should be significantly
smaller for protons and nuclei and become observable only at
much higher energy.

The fundamental idea in this discussion is simple.
It is intuitive that if a large number of source events
contribute to the generation of the observed flux of cosmic rays,
it is a good approximation to consider the source as continuous,
on the contrary if the CR flux is generated 
by only few source events, one can expect observable effects.
When the particle energy $E$ increases,
the space--time volume where the observed CR particles have
their origin shrinks, and therefore the number of source events
that contribute to the flux becomes smaller.
At sufficiently high energy the effects of the discreteness of the sources
(if the sources are indeed discrete) should become manifest.

The discrete nature of the sources can become visible
via effects on the angular and energy distribution of the CR flux.
An excess (or scarcity) of sources in one direction can clearly generate
an anisotropy in the angular distribution of the CR particles, 
and similarly, an excess (scarcity) of near source events can result
in an hardening (softening) of the flux with respect to
the smooth behaviour predicted for a continuous distribution.

There are two important difficulties in constructing a prediction
for the critical energy for source granularity $E^\dagger$.
The first one is that the effects that can reveal this granularity
are model dependent.
The second, more fundamental difficulty is that the observable
effects depend on the real distribution in space and time
of the sources that generate
the cosmic rays, and this distribution is unknown.
Different configurations in space and time of the sources
closer to the Solar System can result in observational effects
that are very different (and of very different sizes).

Keeping in mind these difficulties we will however 
construct some predictions for the energy $E^\dagger$ as a guide to
interpret existing and future observations.
The predictions are based on three elements.
(i) A model for CR propagation in the Galaxy.
(ii) A model for the sources. 
(iii) A criterion to estimate when the granularity effect should become
manifest.

To describe the propagation of CR in the Galaxy we will use the
diffusion model presented in Sec.~\ref{sec:diffusion}.
The model (for one particle type) is entirely defined
by four parameters that can be chosen as: \{$b,H,E^*,\delta$\},
that is the constant that describe energy loss ($|dE/dt| = b \, E^2$),
the vertical size of the CR halo, the critical energy $E^*$ and the exponent
that describes the rigidity dependence of the diffusion coefficient.

To keep the discussion as simple as possible,
we will assume that the source events are all identical,
each generating a CR population $Q_0 ~E^{-\alpha}$.
The CR source spectrum can then be written as the sum:
\begin{equation}
 q(E, \vec{x}, t) = Q_0 ~E^{-\alpha} ~\sum_{k} ~\delta[\vec{x} - \vec{x}_k] ~
 \delta [t - t_k]~
\label{eq:q-discrete}
\end{equation}
with the index $k$ that runs over all source events.
The quantities $\vec{x}_k$ and $t_k$ are the position and time of the
$k$--th event.
It is obviously not possible to predict the space--time
positions of the source events, but one can make assumptions about
the probability distribution for these events. 
We will assume that the source events form on the Galactic plane,
independently from each other with a constant
probability density (per unit time and unit area) $p_s$:
\begin{equation}
p_s (\vec{x}, t) = \frac{n_s}{T_s} \; \delta [z]~.
\label{eq:prob-s}
\end{equation}
In this equation the quantity $T_s^{-1}$
is the frequency of source events in the entire Galaxy
(so if supernova explosions are the dominant CR source,
then $T_s \simeq 50$~years),
and $n_s$ is the probability density per unit area
for one source event in the disk of the Galaxy, taken in the vicinity
of the Solar System.

\begin{figure}[bt]
\begin{center}
\includegraphics[width=11.0cm]{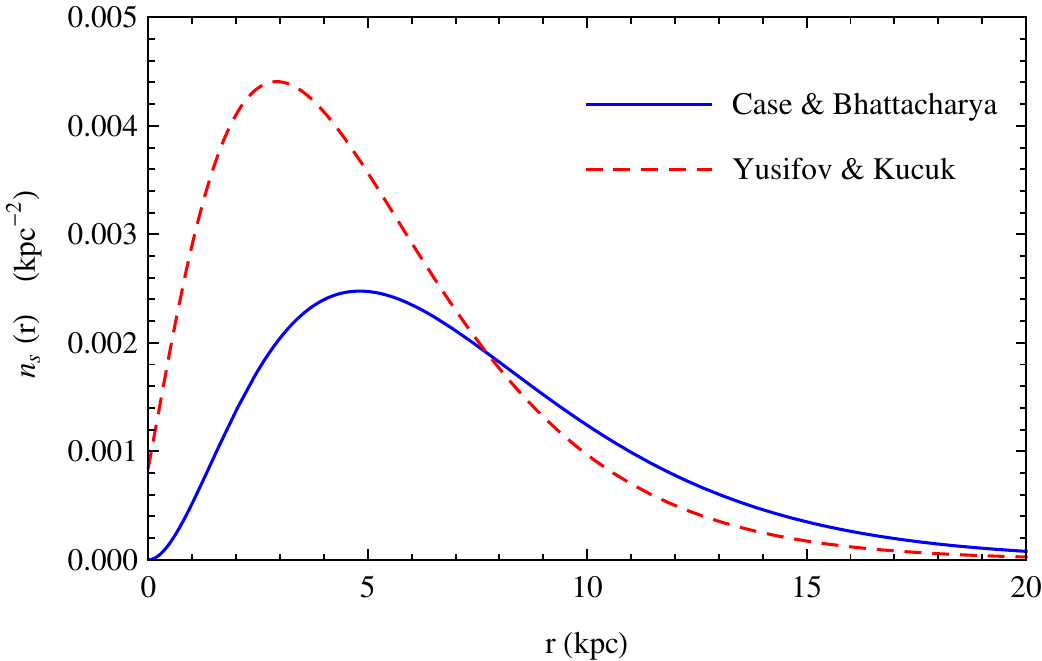}
\end{center}
\caption {\footnotesize
 Fits of the radial distribution of Pulsars
(Yusifov and Kucuk \protect \cite{Yusifov:2004fr})
and supernova remnants (Case and Bhattacharya \protect \cite{Case:1998qg})
in the Mikly Way ($r$ is the distance from the Galactic Center). 
\label{fig:n_sources} }
\end{figure}

To estimate $n_s$ one can use studies of the space distributions
for possible classes of sources such as
Supernovae remnants (SNR) \cite{Case:1998qg} or Pulsars \cite{Yusifov:2004fr}.
These observations show the sources are indeed confined
to narrow layer close to the Galactic equator,
and have a distribution in the plane that can be fitted with a
cylindrically symmetric form $n_s(r)$
(with $r$ the distance from the Galactic Center), that
(by construction) satisfies the normalization condition: 
\begin{equation}
 \int d^2 r ~n_s(r) = 1 ~.
\end{equation}
The fits obtained in \cite{Case:1998qg,Yusifov:2004fr} are shown
in Fig.~\ref{fig:n_sources}. 
The detailed form of $n_s(r)$ can be used as a template for numerical studies
of cosmic ray production. In this work however, we will
simply take $n_s$ as constant: 
$n_s = n_s (r_\odot) \simeq 0.0015$~kpc$^{-2}$,
with the value at the position of the Solar System.
This is a reasonable approximation because at high energy
only sources not too distant from the Earth give significant contributions
to the CR flux (see discussion below), and $n_s (r)$ 
changes rather slowly with $r$.
The (average properties) of the source model are
then determined by 4 parameters: $\{n_s, T_s, Q_0, \alpha\}$.

Combining the propagation and source models,
the description of the CR flux for one particle type
depends on a total of 8 parameters:
$\{b,E^*, H, \delta\}$ for propagation, and
$\{n_s,T_s,Q_0, \alpha\}$ for the source model.
In the discussion below we will consider two parameters 
($b$ and $n_s$) as fixed.
The parameter $b$ (that describes the rate of energy loss) has a value
$b \simeq 0$ for $p$ and $\overline{p}$ while for $e^\mp$ one has 
$b \simeq 5.10 \times 10^{-17}$~(GeV~s)$^{-1}$ 
[see Eq.~(\ref{eq:loss-num})].
The value of $n_s$
(the surface density of the source events normalized to unity
in the entire Galaxy) has been discussed above (see Fig.~\ref{fig:n_sources})
and, in reasonably good approximation, is determined by the size of the
Galactic disk: $n_s \approx (\pi \; R_{\rm disk}^2)^{-1}$.
Other two parameters 
($\alpha$ and $Q_0$) can be determined by the condition that the calculation
matches the observed spectrum (in spectral shape and absolute normalization).
This leaves four parameters as free: $\{T_s,H,E^*,\delta\}$.

Smoothing out the discreteness of the events, the CR source
takes the stationary form:
\begin{equation}
 \left \langle q(E, \vec{x}) \right \rangle
 = \frac{n_s}{T_s} \; Q_0 \; E^{-\alpha} \; \delta [z]
\label{eq:q-cont}
\end{equation}
This source spectrum
(with the identification $q_0 = Q_0 \, n_s /T_s$) is in fact identical
to the source spectrum of Eq.~(\ref{eq:q-diff})
discussed in Sec.~\ref{sec:diffusion}.
The question that we are addressing here is under which
conditions one can expect observable differences between
the flux generated the smoothed out source spectrum of Eq.~(\ref{eq:q-cont})
and the flux generated by an ensemble
of discrete sources as in Eq.~(\ref{eq:q-discrete}).

All models where the combination $Q_0 \, n_s/T_s$
has the same value have the same smoothed out source spectrum,
however this spectrum can be generated by
and ensemble of frequent and faint events
(small $T_s$ and small $Q_0$)
or an ensemble of rare and bright events
(large $T_s$ and large $Q_0$). It is obvious that in the second case the
source granularity is easier to observe.

The simple criterion that we will use to estimate 
the conditions for which the granularity of the sources is manifest,
is the condition that one single source generates a large fraction of the flux.
It is clearly desirable to express this qualitative idea in a more
precise form.
For this purpose one can introduce the concept of the
cumulative flux $\phi_{\rm cum} (N_s, E)$, that is the sum of the largest
$N_s$ contributions to the total flux at energy $E$.

Having a model for the sources and for CR propagation,
the cumulative flux can be calculated for a smoothed out
distribution of the source spectrum, to determine the conditions
for which the brightest source event accounts for one half of the observed flux.
It is intuitive (and it will also be demonstrated in detail below),
that for a fixed model of sources and propagation,
the fraction of the total flux associated to the brighests source
grows with energy, one can therefore solve for $E$ the equation
\begin{equation}
\frac{\phi_{\rm cum} (1, E)}{\phi(E)} = \frac{1}{2}
\label{eq:criterion}
\end{equation}
to calculate an energy $E^\dagger$ that we interpret as
a best estimate of where the source granularity effects shold become manifest.

It should be noted that the criterion that we have constructed for the observability
of the source granularity could very well be too stringent, and it is possible
that evidence for the existence of discrete sources will be be obtained in
situations where dozens (and not just one or very very few)
sources give significant contributions to the CR flux.

In the following we will outline a calculation of the cumulative flux $\phi_{\rm cum} (N_s, E)$.

\subsection{Flux components}
\label{sec:components}

The flux received by one source event at
position $\vec{x}$ and time $t$ can be 
calculated having a model for the spectrum of particles
generated in the event and a model for CR propagation
in the Galaxy. In general one has:
\begin{equation}
 \phi_s (E, \vec{x}, t) = \int dE_i ~Q_s (E_i) ~
 \mathcal{P} (E, \vec{x}_\odot, E_i, \vec{x}, t) ~.
\end{equation}

In the following we will make the (good) approximation that the
Solar system is on the Galactic plane, and assume that all
source events are on this plane, and 
generate identical spectra of particles with a power law spectrum:
$Q_0 \; E^{-\alpha}$. It is then straighforward to obtain
that if energy losses are negligible (that is for $b \simeq 0$),
as is the case for protons and nuclei, the flux from a
source event at a distance $r = |\vec{x}_s - \vec{x}_\odot|$ and
at time $t$ in the past is
\begin{equation}
 \phi_s (E, r, t) =
 \frac{c}{4 \, \pi} ~\frac{Q_0 ~E^{-\alpha}}{H^3} ~
\mathcal{G}_0 \left ( \frac{r} {H}, \frac{t}{T_{\rm esc} (E)} \right ) ~,
\label{eq:gg0}
\end{equation}
where the function $\mathcal{G}_0 (\rho, \tau)$ has the form:
\begin{equation}
\mathcal{G}_0(\rho, \tau) = \frac{1}{2 \, \pi \, \tau} ~e^{-\rho^2/(2 \, \tau)} ~
 g(0,0, \tau)~.
\end{equation}
[with $g$ defined in Eq.~(\ref{eq:gdef})].
The function $\mathcal{G}_0$ satisfies the normalization condition:
\begin{equation}
(2 \, \pi) \int_0^\infty d\rho ~\rho~ \int_0^\infty d\tau ~\mathcal{G}_0 (\rho, \tau) = 1 ~.
\end{equation}

In the case where energy losses are the dominant effect for propagation
(that is the case for high energy electrons and positrons, when
$E \gg E^*$), the spectrum from a single source is:
\begin{equation}
 \phi_s (E, r, t) =
 \frac{c}{4 \, \pi} ~\frac{Q_0 ~E^{-\alpha}}{R_{\rm loss}^3 (E)} ~
 \mathcal{G}_{ \rm loss}
 \left ( \frac{r} {R_{\rm loss}(E) }, \frac{t}{T_{\rm loss} (E)} \right ) ~,
\end{equation}
where the function $\mathcal{G}_{\rm loss} (\rho^\prime, \tau^\prime)$ has the form:
\begin{equation}
\mathcal{G}_{\rm loss} (\rho^\prime, \tau^\prime) = 
 \frac{1}{(2 \, \pi)^{3/2} \; \chi^3 (\tau^\prime)} ~
(1- \tau^\prime)^{\alpha -2} ~ e^{-(\rho^\prime)^2/[2 \,\chi^2 (\tau^\prime)]} 
\label{eq:ggloss}
\end{equation}
with
\begin{equation}
\chi^2(\tau^\prime) = 1 - (1-\tau^\prime)^{1-\delta} 
\end{equation}
[note that $\rho^\prime = r/R_{\rm loss}(E)$ can take values 
in the interval ($0, \infty$)
while $\tau^\prime$ is defined in the interval (0,1)]. 

The function $\mathcal{G}_{\rm loss} (\rho^\prime, \tau^\prime)$ satisfies the normalization
condition:
\begin{equation}
(2 \, \pi) \; \int_0^1 ~d\tau^\prime ~ \int_0^\infty ~d\rho^\prime ~\rho^\prime 
 ~ \mathcal{G}_{\rm loss} (\rho^\prime, \tau^\prime) = k(\alpha, \delta)
\end{equation}
with $k(\alpha, \delta)$ the adimensional quantity given in Eq.~(\ref{eq:kalpha}).

It is important to note that in both cases the shape of the CR flux
generated by a source that is observable at the Earth is {\em not} a simple
power law, even if the source spectrum is a power law.
This is because particles of different energy (or rigidity) propagates
in different ways. This results in an observable spectrum
that evolves with time in normalization and in shape.

For both cases considered the flux generated by 
a single source event at distance $r$ and time $t$
has the scaling form:
\begin{equation}
 \phi_s (r,t)
 = \phi^* ~\mathcal{G} \left ( \frac{r}{R}, \frac{t}{T} \right )
\label{eq:phis-general}
\end{equation}
(where the energy dependence has been left implicit).
In the case of negligible energy loss one has
$R = H$ and $T= T_{\rm esc} (E)$, while
when escape is negligible one has
$R =R_{\rm loss} (E)$ and $t = T_{\rm loss} (E)$.
In Eq.~(\ref{eq:phis-general}) $\phi^*$ is a characteristic flux given by:
\begin{equation}
\phi^* = \frac{c}{4 \, \pi} ~\frac{Q_s(E)}{R^3} 
\label{eq:phistar-general}
\end{equation}
with $Q_s(E)$ the emission from the source.

Lines of constant value for the contributions of a source event
(for some values of the ratio $\phi_s/\phi^*$)
in the plane $\{\rho, \tau\}$
are shown in Fig.~\ref{fig:dens1a}
for the case of negligible energy loss, and in
Fig.~\ref{fig:dens2a} for the case of dominant energy loss.


The contributions from sources at a fixed distance $r$
depends on the time $t$.
For a (arbitrary) fixed value of $E$, the flux received from one source
initially grows with time, reaching its maximum value at the time
$t \simeq r^2/[2 \, D(E)]$, and then decreases slowly.
A contribution of a certain size $\phi_s$ can therefore be obtained
from events that have a distance $r < r_{\rm max} (\phi_s)$. For each
$r < r_{\rm max}$ there are two solutions for the age that give the
same contribution $\phi_s$. The two solutions correspond
to the cases where the flux from the sources is growing or decreasing with time.

A simple but very important point that for the model of diffusive propagation
that we are discussing here, the contribution to the flux of each source event
has associated a dipole momentum in the direction of the source
with a size that can be calculated using the general expression of Eq.~(\ref{eq:dipole}).

The vanishing dipole momentum of the total flux for the
smoothed out source spectrum is the effect of the cancellation
of the dipoles of the different source events that are located
symmetrically around the Solar system, but the contribution of the closest source events
will not balance exactly, because of stochastic effects. If few sources
contribute to the flux one can expect large anisotropy effects.

\subsection{Space--time distribution of the CR source} 
\label{sec:space-time}
It is instructive to study the size and shape of the space--time volume
where the cosmic rays observed at the Earth are generated.
In general, the space time distribution of the CR sources can be
calculated from a knowledge of the source spectrum and the propagation
function as:
\begin{equation}
 \frac{d\phi(E, \vec{x}_i, t)} {d^3 x_i \, dt} =
 \int dE_i ~
 q(E_i, \vec{x}_i, t) ~
 \mathcal {P} (E, \vec{x}_\odot, E_i, \vec{x}_i, t)
\end{equation}
(where $t=0$ is the present, and $\vec{x}_\odot$ is the position of the
Solar System).

\begin{figure}[bt]
\begin{center}
\includegraphics[width=11.0cm]{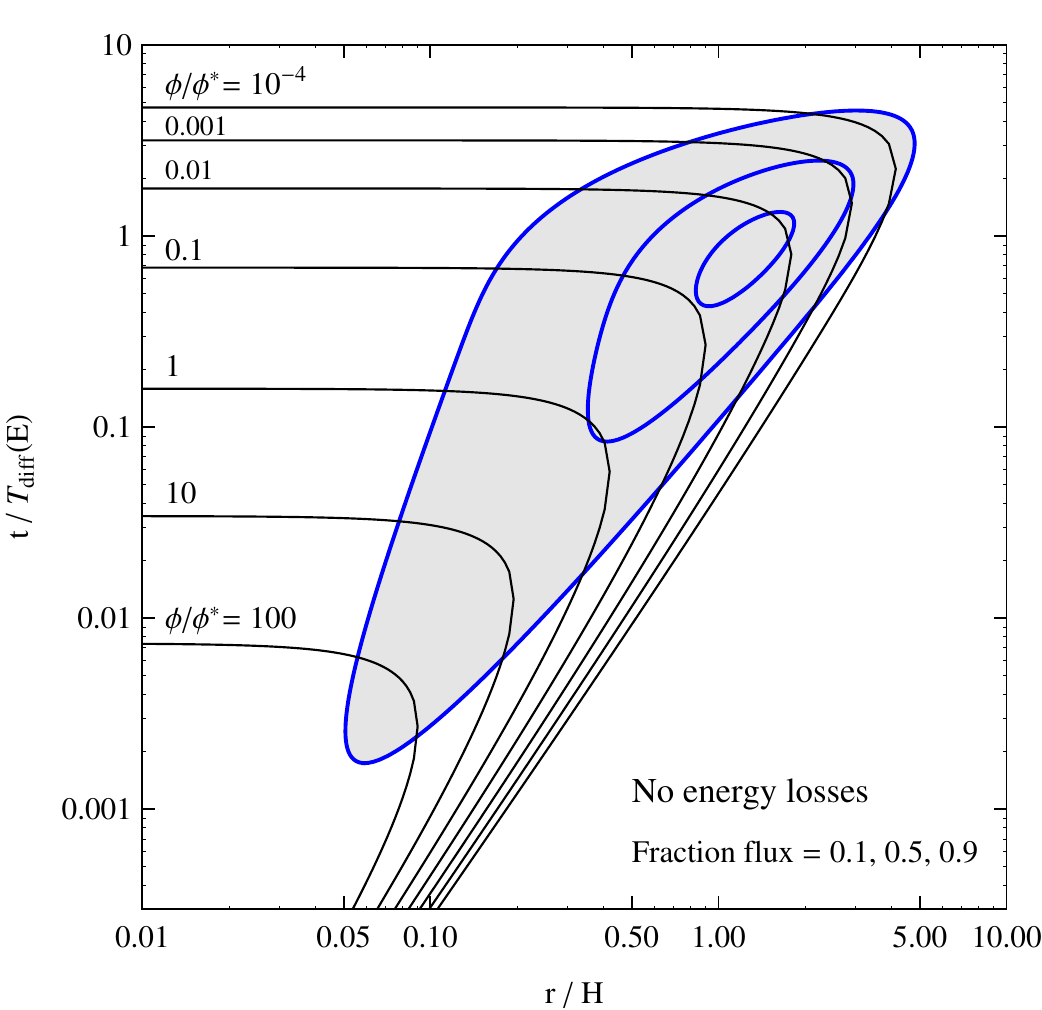}
\end{center}
\caption {\footnotesize
Space--time distribution of the CR flux
(energy losses are considered negligible).
The thick closed lines in and around the shaded area
are defined by the equation $d\phi/(d\ln r \, d\ln t) = $~const,
and chosen so that the integral in the region inside the line
corresponds to a fraction 0.1, 0.5 and 0.9 of the total flux.
The thinner lines are defined by the equation
$\phi_s (r, t)/\phi^* = f$, with $f = 10^{-4}$,
$10^{-3}$, $10^{-2}$, $10^{-1}$, 1, 10 and $10^2$.
\label{fig:dens1a} }
\end{figure}

\begin{figure}[bt]
\begin{center}
\includegraphics[width=7.0cm]{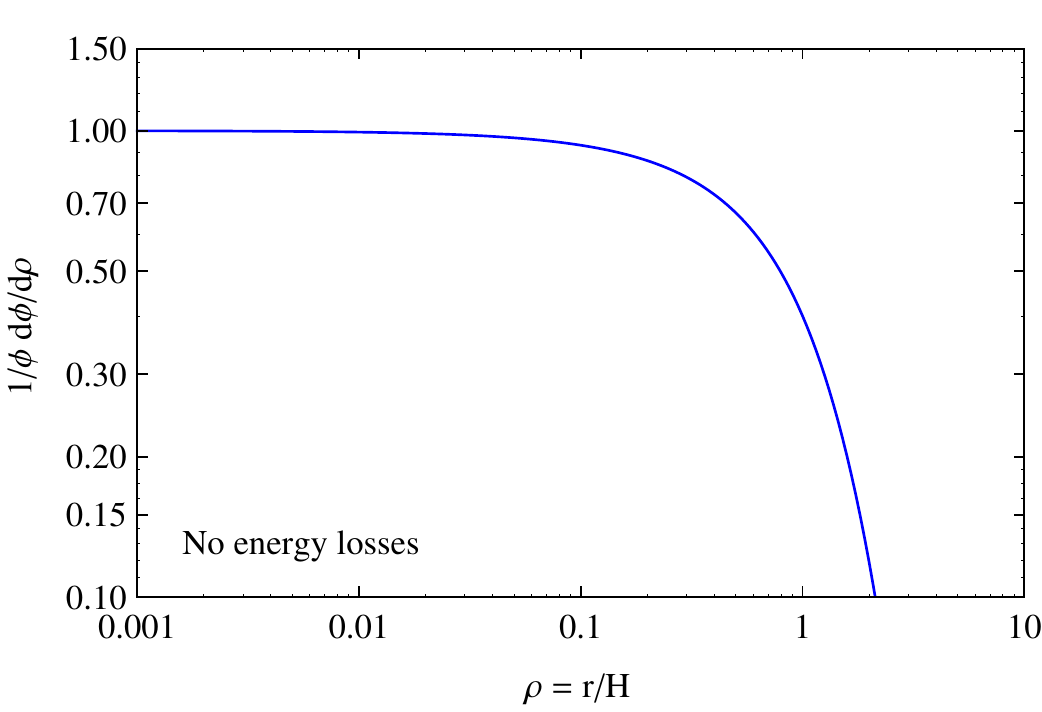}
~~~~
\includegraphics[width=7.0cm]{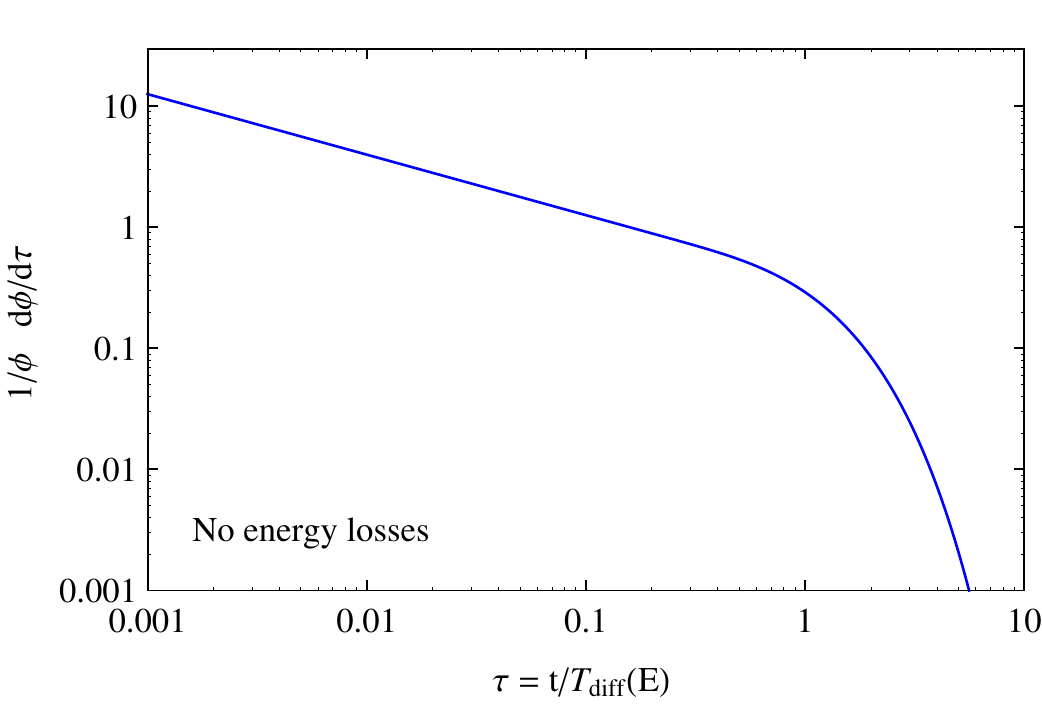}
\end{center}
\caption {\footnotesize
Space and time distribution of the CR sources.
(energy losses are considered as negligible).
Top panel: plot of $1/\phi ~d\phi/d\rho$ versus $\rho = r/H$. 
Bottom panel: plot of $1/\phi ~d\phi/d\tau$
versus $\tau = t/T_{\rm esc} (E)$. 
\label{fig:radius1} }
\end{figure}

For the model we are discussing here,
and a smoothed out source spectrum,
the space--time volume takes a simple scaling form
in the cases of negligible energy loss and negligible escape probability.
In the first case one finds:
\begin{equation}
\left . \frac{d\phi(E, r_i, t)} {d^2r \, dt} \right |_{\rm no~losses} = 
\frac{\phi(E)}{H^2 \; T_{\rm esc} (E)} ~\mathcal{G}_0 \left (
\frac{r}{H}, \frac{t}{T_{\rm esc}(E)}
\right )
\end{equation}
where $r$ is the distance of the source event,
and the function $\mathcal{G}_0 (\rho, \tau)$ is given in Eq.~(\ref{eq:gg0}).
A graphic representation of the function
$\rho^2 \, \tau \, \mathcal{G}_0(\rho, \tau)$
[that is proportional to the distribution $d\phi/(d \ln r \; d\ln t)$] is shown,
in the form of a contour plot in Fig.~\ref{fig:dens1a}.
The $\rho$ and $\tau$ distributions of the source
can be obtained integrating over the other variable,
and are shown in Fig.~\ref{fig:radius1}. 

These results show that
CR observed at the Earth have their origin
in a time interval of order $T_{\rm esc} (E)$
(with the contributions of older emission suppressed exponentially),
and in a space region that is a approximately
a disk centered on the position of the solar system
with a radius of order $H$ (and again the contributions of more distant sources
are suppressed exponentially).

The time interval in which the observed CR are produced becomes shorter
with increasing energy, because $T_{\rm esc}(E)$
decreases $\propto E^{-\delta}$, on the other hand the 
space region where the CR are produced is energy independent.
This is the result of a cancellation:
particles of higher energy have a shorter Galactic residence time,
but also have a larger diffuse coefficient,
the result is a propagation distance that is independent
from energy.

For the case where energy losses are the dominant effect in propagation,
the space--time distribution of the source takes the form:
\begin{equation}
\left . \frac{d\phi(E, r, t)} {d^2r \, dt} \right |_{\rm no~escape}
= \frac{\phi(E)}{R_{\rm loss}^2(E) \; T_{\rm loss} (E)} \;
\frac{1}{k(\alpha,\delta)}
 ~\mathcal{G}_{\rm loss}
\left (
\frac{r}{R_{\rm loss}(E) }, \frac{t}{T_{\rm loss}(E)} ~.
\right )
\end{equation}
where the function $\mathcal{G}_{\rm loss} (\rho^\prime, \tau^\prime)$
is given in Eq.~(\ref{eq:ggloss}). 

\begin{figure}[bt]
\begin{center}
\includegraphics[width=11.0cm]{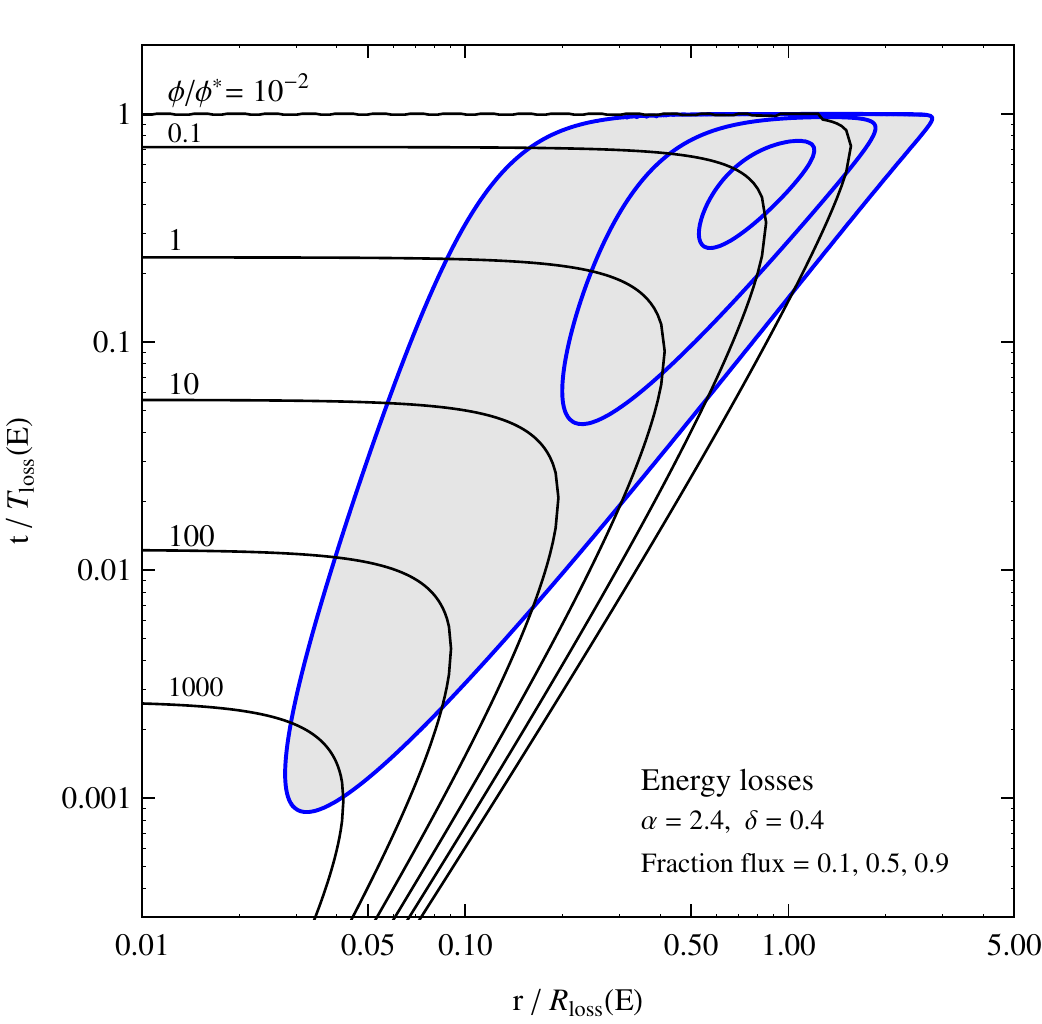}
\end{center}
\caption {\footnotesize
Space--time distribution of the CR flux
(energy losses are considered as dominant).
See Fig.~\ref{fig:dens1a} for the meaning
of the different lines.
The calculation is performed assuming that the source
spectrum is a power law with exponent $\alpha =2.4$,
and the diffusion coefficient has the energy
dependence $D_0 \; E^\delta$ with $\delta =0.4$.
\label{fig:dens2a} }
\end{figure}

\begin{figure}[bt]
\begin{center}
\includegraphics[width=7.0cm]{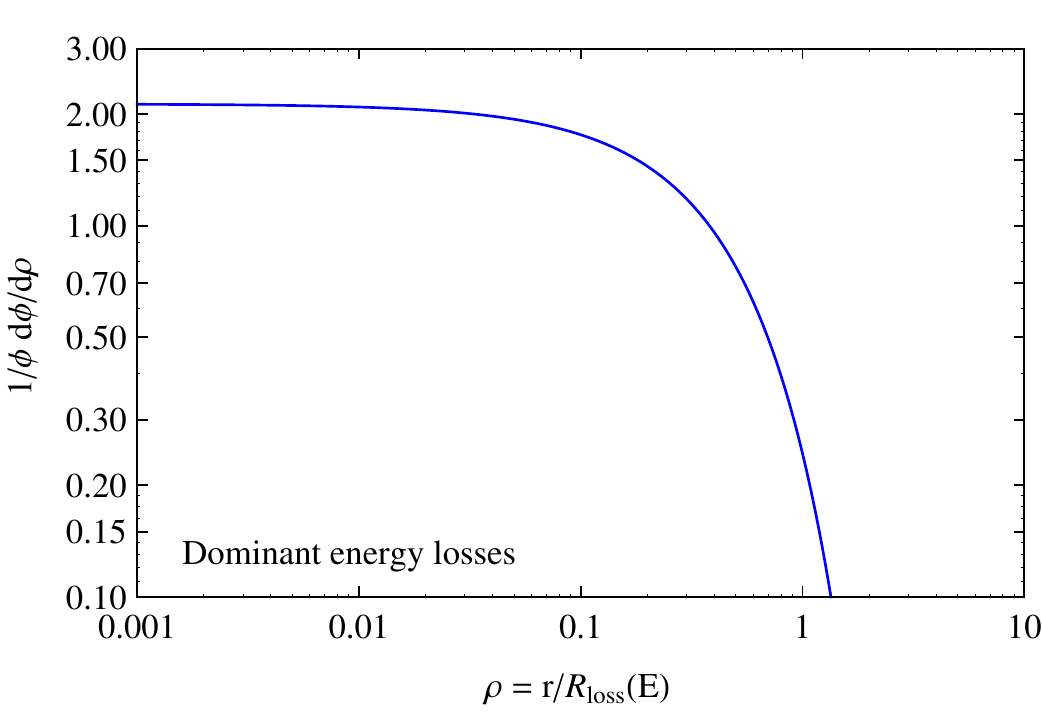}
~~~~
\includegraphics[width=7.0cm]{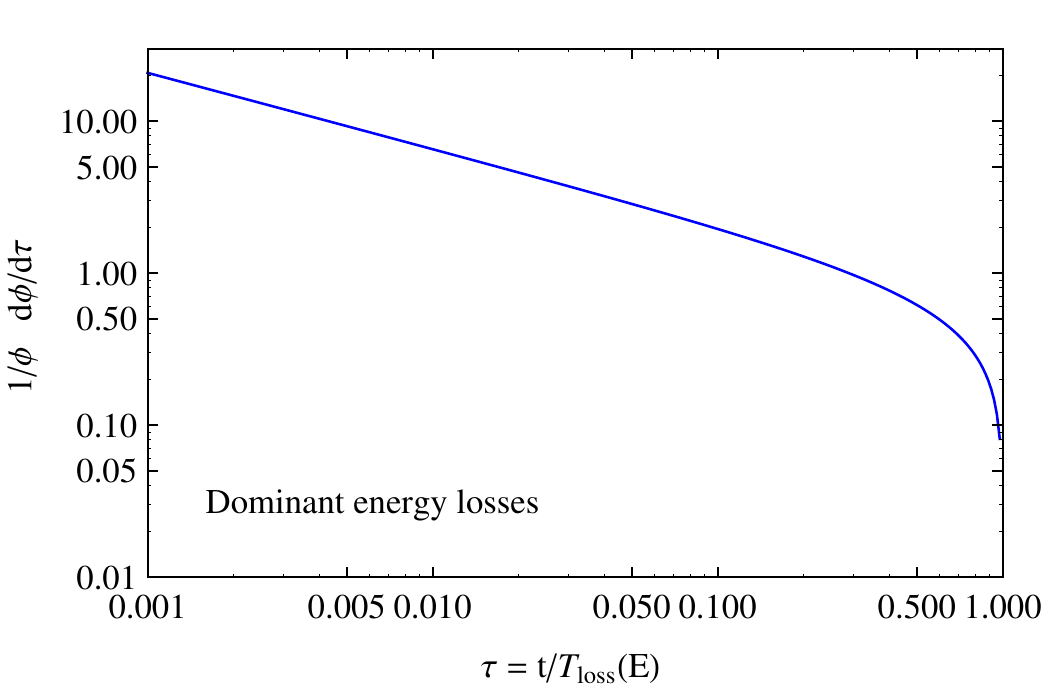}
\end{center}
\caption {\footnotesize
Space and time distribution of the CR sources.
(energy losses are dominant).
Top panel: plot of $1/\phi ~d\phi/d\rho^\prime$ versus $\rho^\prime = r/T_{\rm max}(E)$. 
Bottom panel: plot of $1/\phi ~d\phi/d\tau^\prime$
versus $\tau^\prime = t/T_{\rm loss} (E)$. 
 \label{fig:radius2} }
\end{figure}

The function $(\rho^\prime)^2 \, \tau^\prime \,
\mathcal{G}_{\rm loss} (\rho^\prime, \tau^\prime)$
is shown in the form of a contour plot in Fig.~\ref{fig:dens2a}, and
the $\rho^\prime$ and $\tau^\prime$ distributions
(obtained integrating over the other variable)
are shown in Fig.~\ref{fig:radius2}.

These results demonstrate that in this case
CR observed at the Earth have their origin
in a time interval of order $T_{\rm loss} (E)$
and in space region that is approximately a disk of radius $R_{\rm loss} (E)$,
with the contributions of older and more distant sources suppressed exponentially.

The time $T_{\rm loss}$ decreases with energy
$\propto E^{-1}$, and the radius $R_{\rm loss} (E)$
decreases $\propto E^{-(1 - \delta)/2}$ (because the growth with $E$ 
of the diffusion coefficient is not sufficiently rapid to compensate
for the shorter propagation time). The space--time volume of the CR sources
therefore shrinks rapidly with energy $\propto E^{-(2 -\delta)}$,
much more rapidly than in the case where the energy loss is negligible.

\subsection{Cumulative flux} 
\label{sec:cumulative}
Assuming that the CR flux is formed by the sum of the contributions
of discrete events, and considering a smoothed out source
(so that the quantities $N_s$ and $\phi_s$, that is the number of sources
and the size of source contribution can be considered as continuous variables)
it is straightforward to compute the differential distributions
$dN_s/d\phi_s$ that gives the number of sources
with flux in the interval $[\phi_s,\phi_s + d\phi_s]$, and
$d\phi/d\phi_s$ that gives the flux generated by contributions
in the same size interval.
For the model discussed here,
these quantities can be calculated performing the integrals:
\begin{equation}
 \frac{dN_s}{d\phi_s} =
 \frac{n_s}{T_s} ~
 \int ~d^2 r \int ~dt
 ~\delta [\phi_s - \phi_s(r,t)]
\end{equation}
\begin{equation}
 \frac{d\phi}{d\phi_s} = \frac{n_s}{T_s}
 ~\int ~d^2 r \int ~dt 
 ~ \phi_s ~\delta [\phi_s - \phi_s(r,t)] ~.
\end{equation}
(where we have left the energy dependence implicit).
Using the results of Sec.~\ref{sec:components}
and the general form of the flux from each source
of Eq.~(\ref{eq:phis-general}) one finds that the 
distributions have the simple scaling forms:
\begin{equation}
 \frac{dN_s}{dx} = N^* ~\mathcal{A}\left ( x \right )
\label{eq:dnsdx}
\end{equation}
\begin{equation}
 \frac{d\phi}{dx} = N^*~\phi^* ~\mathcal{B}\left ( x \right )
\label{eq:dphidx}
\end{equation}
where $x= \phi_s/\phi^*$ with $\phi^*$ is
the characteristic flux given in Eq.~(\ref{eq:phistar-general}).
The energy dependent quantity $N^*$ is given by:
\begin{equation}
N^* = \frac{n_s}{T_s} \; R^2 \; T ~.
\label{eq:nstar-general}
\end{equation}
and in first approximation gives the total number of sources
that contribute to the CR flux.

The total flux can be expressed in terms of $\phi^*$ and $N^*$ as:
\begin{equation}
\phi(E) = k ~N^* (E) ~\phi^* (E)
\end{equation}
where $k$ is an energy independent constant given by the integral
\begin{equation}
 k = (2 \, \pi) ~\int_0^{\tau_{\rm max}} ~\int_0^\infty d\rho ~\rho~\mathcal{G}
 (\rho, \tau) ~.
\label{eq:k-general}
\end{equation}
The quantity $\tau_{\rm max}$ is the maximum value of
$\tau$ and has value $\tau_{\rm max} \to \infty$
in the case of negligible energy loss, and
$\tau_{\rm max} = 1$ when energy loss is dominant in propagation.
As shown above, the constant $k$ is unity in the case of negligible
energy loss, and for the case where energy losses are dominant is given
by Eq.~(\ref{eq:kalpha}). 
The functions $\mathcal{A}(x)$ and $\mathcal{B}(x)$
can be calculated as: 
\begin{equation}
 \mathcal{A}(x)
 = (2 \, \pi) \; \int d\rho~\rho~\int_0^{\tau_{\rm max}} d\tau
 ~\delta [x -\mathcal{G}(\rho, \tau)]
\label{eq:A-diff}
\end{equation}
\begin{equation}
 \mathcal{B}(x)
 = (2 \, \pi) \; \int d\rho~\rho~\int_0^{\tau_{\rm max}} d\tau
 ~x ~\delta [x -\mathcal{G}(\rho, \tau)] ~.
\label{eq:B-diff}
\end{equation}

\begin{figure}[bt]
\begin{center}
\includegraphics[width=8.0cm]{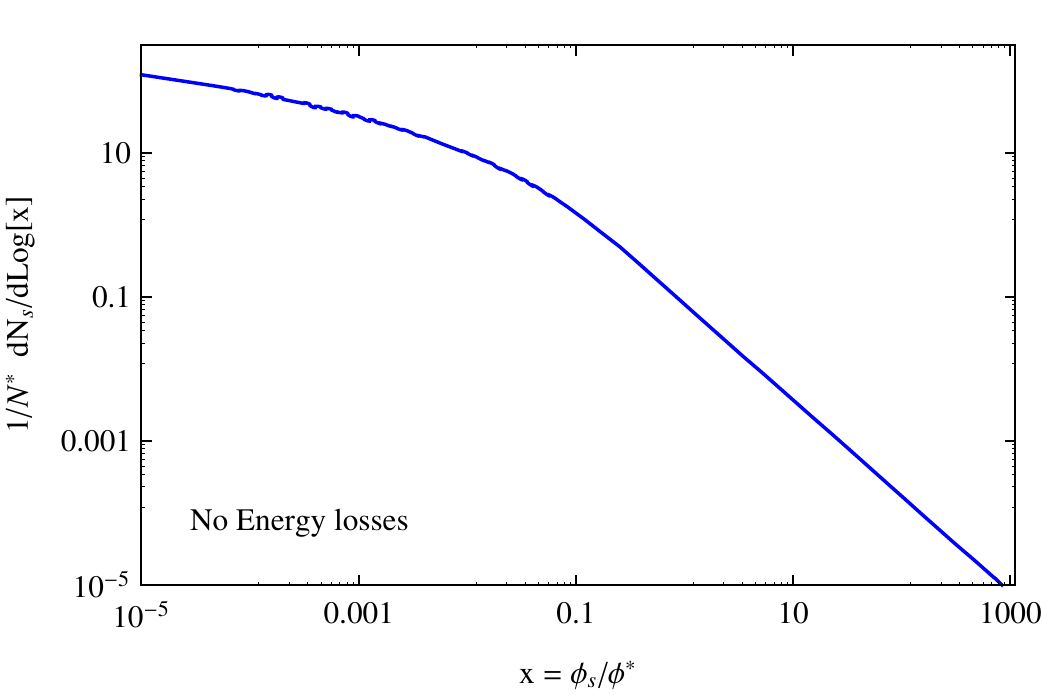}
\end{center}
~~~
\begin{center}
\includegraphics[width=8.0cm]{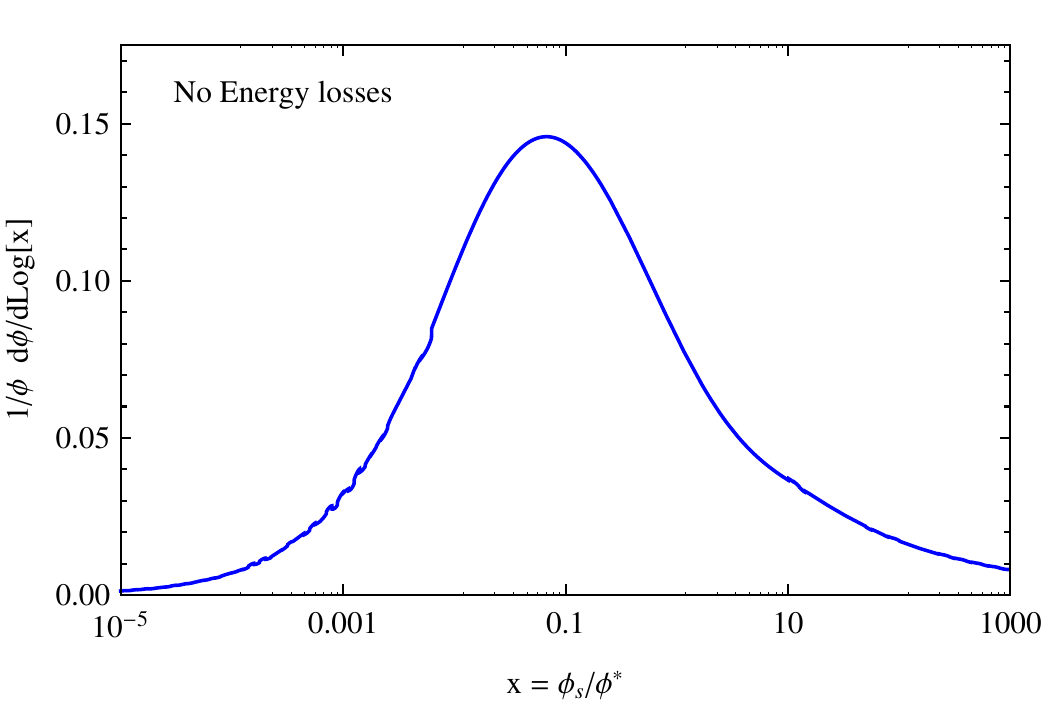}
\end{center}
\caption {\footnotesize
 The left (right)
 panel shows the shape of the distribution $dN_s/d\phi_s$
 ($d\phi/d\phi_s$) as a function of $\phi_s$.
 calculated assuming that energy losses are negligible.
 The distributions are shown in the energy independenty scaling
 form, plotting the quantity $1/N^* ~ dN_s/d\ln x$
 ($1/\phi ~ d\phi/d\ln x$ in the bottom panel) where
 $x = \phi_s/\phi^*$ and the quantities
 $N^*$ and $\phi^*$ are given in 
 Eqs.~(\ref{eq:nstar-general}) and~(\ref{eq:phistar-general}).
 with $R = H$ and $T = T_{\rm esc} (E)$.
\label{fig:dns_noloss} }
\end{figure}

In the case of negligible energy loss, the functions
$\mathcal{A}_0 (x)$ and $\mathcal{B}_0 (x)$
(that determine the shape of the distributions
$dN_s/d\phi_s$ and $d\phi/d\phi_s$) 
have a universal
shape that is independent from the values of the parameters
in the model. The functions are shown in Fig.~\ref{fig:dns_noloss}.
Inspecting the figure one can see that the observed flux is formed by
components that have a very broad distributions of relative size,
that spans approximately six orders of magnitudes. 

\begin{figure}[bt]
\begin{center}
\includegraphics[width=8.0cm]{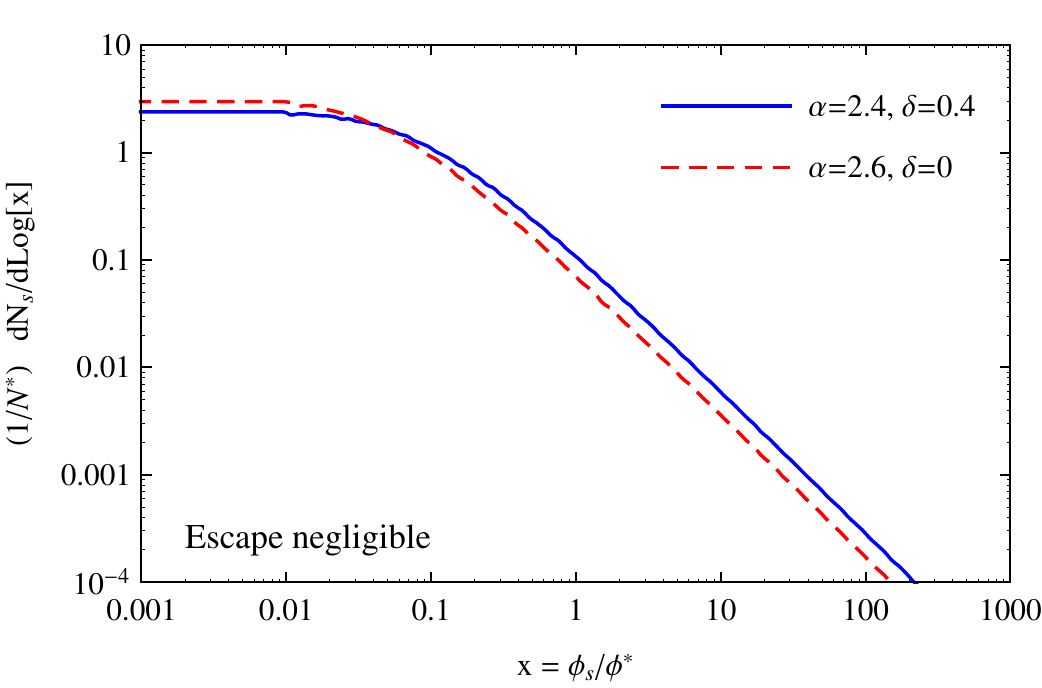}
\end{center}
~~~
\begin{center}
\includegraphics[width=8.0cm]{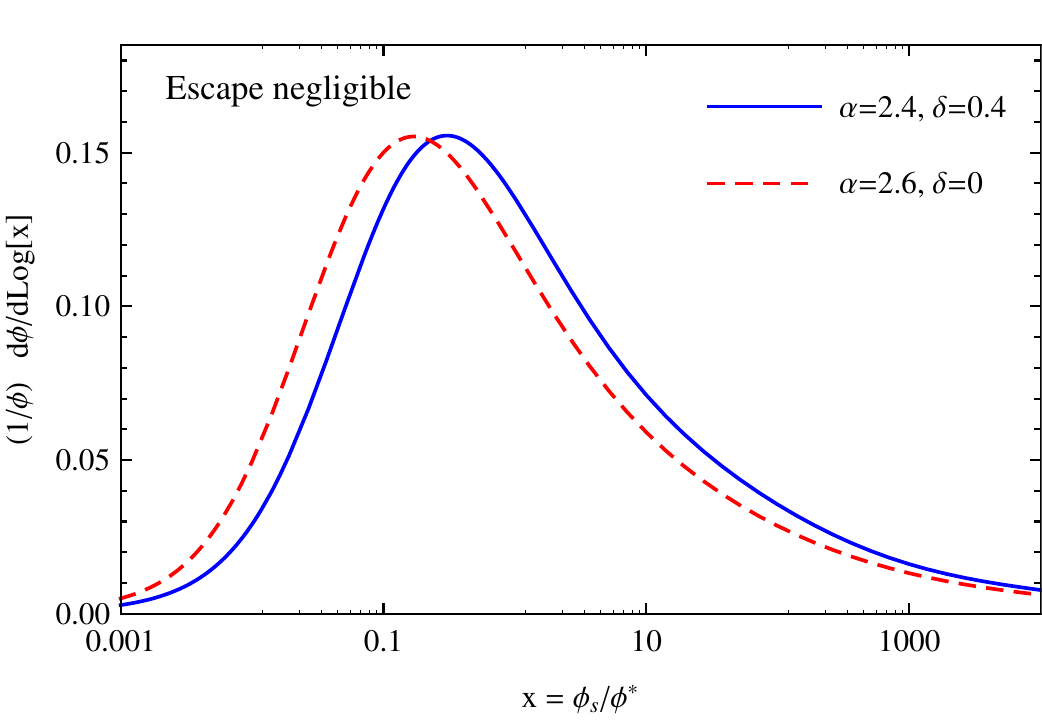}
\end{center}
\caption {\footnotesize
 The left (right)
 panel shows the shape of the distribution $dN_s/d\phi_s$
 ($d\phi/d\phi_s$) as a function of $\phi_s$.
 calculated assuming that energy losses are dominant
 for CR propagation (escape is negigible).
 The distributions are shown in the energy independenty scaling
 form, plotting the quantity $1/N^* ~ dN_s/d\ln x$
 ($1/\phi ~ d\phi/d\ln x$ in the bottom panel) where
 $x = \phi_s/\phi^*$ and the quantities
 $N^*$ and $\phi^*$ are given in 
 Eqs.~(\ref{eq:nstar-general}) and~(\ref{eq:phistar-general}).
 with $R = R_{\rm loss} (E)$ and $T = T_{\rm loss} (E)$.
 The shape of the distributions depends on the exponents
 $\alpha$ and $\delta$, and the curved describe
 the cases \{$\alpha = 2.6$, $\delta=0.4$\}
 and \{$\alpha = 2.6$, $\delta=0.0$\}.
 \label{fig:dns_loss} }
\end{figure}

In the case where energy loss is dominant, the shape of the functions
$\mathcal{A}_{\rm loss} (x)$ and $\mathcal{B}_{\rm loss} (x)$ is determined
by the values of the exponents $\alpha$ and $\delta$.
Two examples are shown in Fig.~\ref{fig:dns_loss}.
The two curves (in both panels) are calculated for
the combinations ($\alpha = 2.4$, $\delta = 0.4$) and
($\alpha = 2.6$, $\delta = 0$). Note that in both cases
the observed flux (for a smoothed out source spectrum)
has the same spectral index: $\gamma \simeq \alpha + (1+ \delta)/2 \simeq 3.1$.
Also in this case one can see that the CR flux is formed by
components that have very different size

It is intersting to discuss how the quantity $N^* (E)$,
defined in Eq.~(\ref{eq:nstar-general}) depends on the parameters of the model.
In the case of negligible energy loss, that is applicable to
protons, one has:
\begin{equation}
 N_p^* = \frac{n_s}{T_s} \; H^2 \; T_{\rm esc} (E) =
 \frac{n_s}{T_s} \; \frac{H^2}{b} ~ (E^*)^{-(1-\delta)} ~E^{-\delta} 
\end{equation}
(in the second equality the escape
time has been expressed as function of the critical energy $E^*$).
The number of sources that generate the flux of protons
(and other primary nuclei)
decreases with energy $\propto E^{-\delta}$, and has
a value that depends on the model parameters as:
$N_p^* \propto T_s^{-1} \; H^2 \; (E^*)^{-(1-\delta)}$.
The quantity is proportional to the frequency of source events ($T_s^{-1}$)
and it becomes smaller when $E^*$ increases (that is when the CR
residence time becomes smaller).
A numerical example (for $\delta = 1/3$) is:
\begin{equation}
 N_p^* (E) \simeq 2240
 ~\left [ \frac {T_s} {50 ~{\rm yr}} \right ]^{-1}
 ~\left [ \frac{H} {5 ~{\rm kpc}} \right ]^2
 ~\left [ \frac{E^*} {3 ~{\rm GeV}} \right ]^{-2/3}
 ~\left ( \frac{E} {{\rm PeV}} \right )^{-1/3} ~.
\label{eq:n-num1}
\end{equation}
(note that $E$ is measured in PeV).
One can see that (for the propagation model considered here)
if supernova explosions are the main CR source,
the number of events that contribute to the proton flux
remains rather large also for energies well above the ``Knee''
(at $E \simeq 3$~PeV). This statement remains true also
if the critical energy $E^*$ has a value close to 1~TeV
(in this case the estimate in Eq.~(\ref{eq:n-num1}) is reduced 
to approximately 50).

For the case where energy losses are the dominant effect in propagation,
that is relevant for $e^\mp$ at high energy, one has
\begin{equation}
 N_e^* = \frac{n_s}{T_s} \; R_{\rm loss}^2(E) \; T_{\rm loss} (E) =
 \frac{n_s}{T_s} \; \frac{H^2}{b \, (1-\delta)} ~ (E^*)^{(1-\delta)} ~E^{-(2-\delta)} 
\end{equation}
(again in the second equality we have expressed $N_e^*$
in terms of the critical energy $E^*$).
In this case, the number of sources
that contribute to the flux 
decreases rapidly with energy ($\propto E^{-(2-\delta)}$),
reflecting the fact that both $R_{\rm loss}$ and $T_{\rm loss}$ decrease
with $E$. The dependence of $N_e^*$ on the parameters
of the model is 
$N_e^* \propto T_s^{-1} \; H^2 \; (E^*)^{(1-\delta)}$.
Note that in this case the estimate of $N_e^*$ increases
with $E^*$. This reflects the fact that with increasing $E^*$
the CR residence time become shorter, but the propagation radius grows,
and this second effect is dominant.
A numerical example (for $\delta = 1/3$) is:
\begin{equation}
 N_e^* (E) \simeq 9.6
 ~\left [ \frac {T_s} {50 ~{\rm yr}} \right ]^{-1}
 ~\left [ \frac{H} {5 ~{\rm kpc}} \right ]^2
 ~\left [ \frac{E^*} {3 ~{\rm GeV}} \right ]^{2/3}
 ~\left ( \frac{E} {\rm TeV} \right )^{-5/3} 
\label{eq:n-num2}
\end{equation}
(note that $E$ is measured in TeV).
 This estimate shows that if supernova explosions are the sources of
 electrons, and if the critical energy $E^*$ is in the GeV range, 
(and the CR propagation model used here is reasonably correct)
then, only few sources contribute to the flux for $E \gtrsim 1$~TeV.

The calculation of $N^*_{e,p}$ allows already a first order estimate
of when the source granularity effects shoud become visible,
using for example the equation $N^*_{e,p} (E^\dagger) \simeq 1$ to
calculate the energy where the effects should be large.
In fact (as discussed in the following) this simple
argument yields results that are in good approximation correct
for protons, and overstimate the energy 
$E^\dagger$ by a factor 3--5 for $e^\mp$.
It is however desirable to construct a more robust
argument for the estimate of $E^\dagger$, as discussed below.


The results 
of Eqs.~(\ref{eq:dnsdx}) and~(\ref{eq:dphidx})
for the differential distributions $dN_s/d\phi_s$ and
 $d\phi/d\phi_s$ imply that the quantities $N_s(\phi_{s, {\rm min}})$
[the number of sources that generate a flux larger
than $\phi_{s, {\rm min}}$] and
$\phi_{\rm cum} (\phi_{s, {\rm min}})$ [the flux formed
by components that have flux larger than $\phi_{s, {\rm min}}$]
also have a scaling form:
\begin{equation}
 N_s (\phi_{s, {\rm min}}) = N^* ~\mathcal{A}_{\rm int} \left (
 \frac{\phi_{s, {\rm min}}} {\phi^*} \right )
\end{equation}
and 
\begin{equation}
 \phi_{\rm cum} (\phi_{s, {\rm min}})
 = \phi ~\mathcal{B}_{\rm int} \left (
 \frac{\phi_{s, {\rm min}}} {\phi^*} \right )
\end{equation}
where the functions
$\mathcal{A}_{\rm int}(x)$ and $\mathcal{B}_{\rm int}(x)$ are given by:
\begin{equation}
\mathcal{A}_{\rm int} (x) = \int_x^\infty dx^\prime ~\mathcal{A} (x^\prime)
\end{equation}
\begin{equation}
 \mathcal{B}_{\rm int} (x) = \frac{1}{k}
 ~\int_x^\infty dx^\prime ~\mathcal{B} (x^\prime)
\end{equation}
[with $k$ given by Eq.~(\ref{eq:k-general}).
 Combining the last two equations,
 it is straightforward to express the cumulative flux
 $\phi_{\rm cum}$ as a function of the $N_s$ (the number of largest sources):
\begin{equation}
\phi_{\rm cum} (N_s) = \phi ~\mathcal{C} \left (\frac{N_s}{N^*} \right )
\label{eq:phicum-scaling}
\end{equation}
where the function $\mathcal{C} (m) $ is:
\begin{equation}
\mathcal{C} = \mathcal{B}_{\rm int} [ \mathcal {A}^{-1}_{\rm int} (m)] ~.
\end{equation}

\begin{figure}[bt]
\begin{center}
\includegraphics[width=11.0cm]{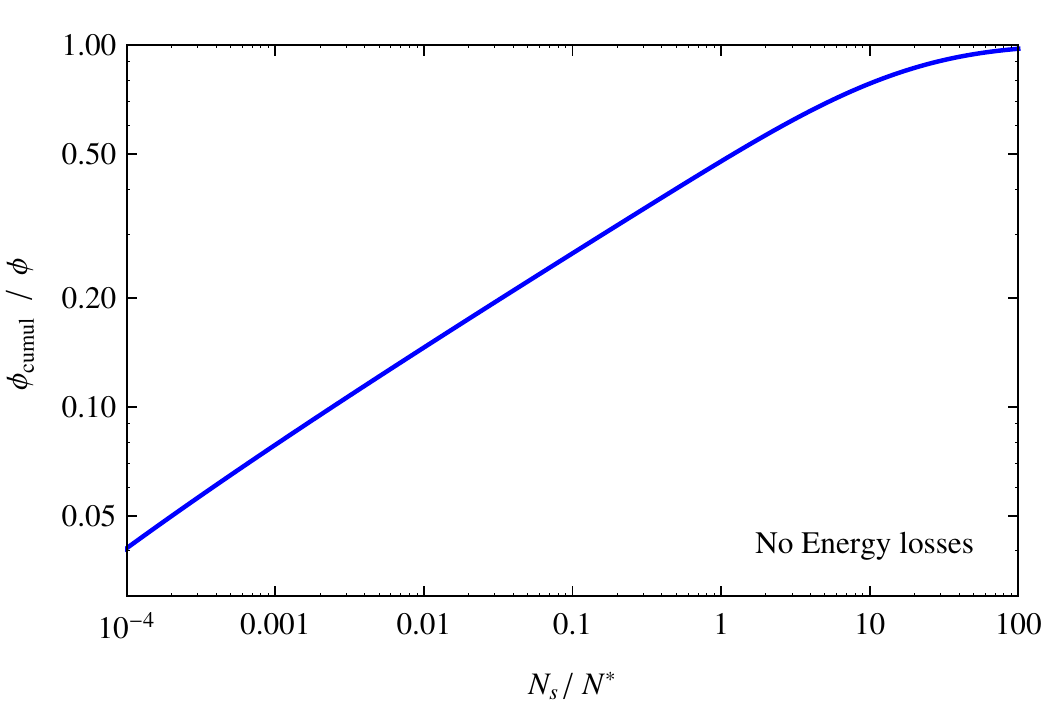}
\end{center}
\caption {\footnotesize
 Cumulative flux $\phi_{\rm cum}$ as a function of the number of
 source events $N_s$ calculated assuming that energy losses
 are negligible. The relation is shown in the energy independent
 form $\phi_{\rm cum} (N_s) /\phi$ versus $N_s/N^*$ where $N^*$ is given in
 Eq.~(\ref{eq:nstar-general})
 with $R = H$ and $T = T_{\rm esc} (E)$]. 
\label{fig:cumul0} }
\end{figure}

\begin{figure}[bt]
\begin{center}
\includegraphics[width=11.0cm]{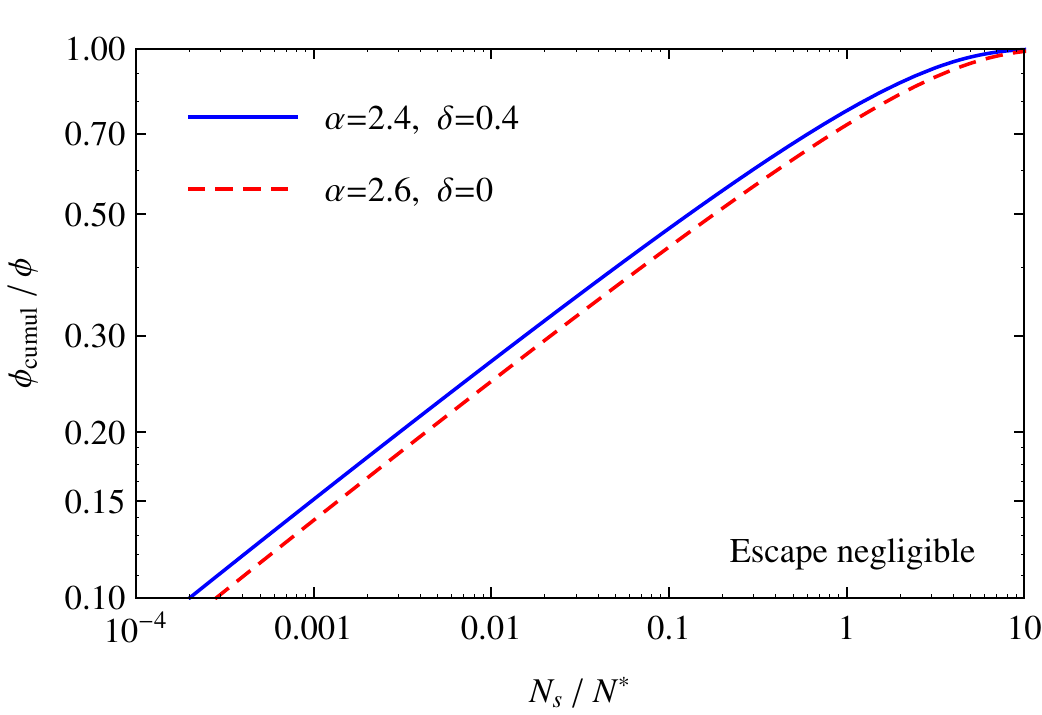}
\end{center}
\caption {\footnotesize
 As in Fig.~\ref{fig:cumul0}, but the 
 flux calculated assuming that energy losses are
 the dominant effect in propagation.
 The quantities $N^*$ is defined in Eq.~(\ref{eq:nstar-general})
 with $R = R_{\rm loss} (E)$ and $T = T_{\rm loss} (E)$.
 The shape of the curve depends on the exponents
 $\alpha$ and $\delta$. The two lines in the figure describe
 the cases \{$\alpha = 2.6$, $\delta=0.4$\}
 and \{$\alpha = 2.6$, $\delta=0.0$\}.
\label{fig:cumul_loss} }
\end{figure}

The relation between $\phi_{\rm cum}$ and the $N_s$ 
is shown for the cases of negligible energy loss and
negligible escape in Fig.~\ref{fig:cumul0} and~\ref{fig:cumul_loss}.
As already discussed, the shape of the relation between the ratios
$\phi_{\rm cum} (N_s)/\phi$ and $N_s/N^*$ is universal (that is independent
from the values of the parameters of the model) for the case of
negligible energy losses, and depends on the exponents
$\alpha$ and $\delta$ when energy losses are dominant.

Having in hand an explicit expression for the cumulative flux as a function of the
number of sources [see Eq.~(\ref{eq:phicum-scaling})],
the estimate of the energy where the source granularity effects should
become manifest according to the criterion of Eq.~(\ref{eq:criterion})
can be stated as:
\begin{equation}
 N^* (E) =
 \frac{n_s} {T_s} ~R^2(E) \; T(E) \simeq \frac{1}{\mathcal{C}^{-1} (1/2)} ~.
\label{eq:criterion1}
\end{equation}
It should be noted that this equation has the same structure
as the ``naive'' criterion $N^* (E) =1 $ (simply replacing
unity with the number $\mathcal{C}^{-1} (1/2)$),
but is now derived in a more rigorous way.

Using the calculations of the functions $\mathcal{C}_0 (m)$ and
$\mathcal{C}_{\rm loss} (m)$ shown in Figs.~\ref{fig:cumul0} and~\ref{fig:cumul_loss})
that give
$1/\mathcal{C}_0^{-1} (1/2) =1.23$ and
$1/\mathcal{C}_{\rm loss} ^{-1} (1/2) =7.76$ (5.54)
for the choice of parameters
\{$\alpha = 2.6$, $\delta=0.4$\} (\{$\alpha = 2.6$, $\delta=0.0$\}),
one can obtain quantitative estimates for $E^\dagger$.

In the case of protons (negligible energy losses)
the granularity energy has a very high value. For $\delta = 1/3$
one finds:
\begin{equation}
 \left ( E^\dagger \right )_{p}
 \simeq 5.4 \times 10^{19}~{\rm eV} ~
 ~\left [ \frac{T_s} {50 ~{\rm yr}} \right ]^{-35}
 ~\left [ \frac{H} {5 ~{\rm kpc}} \right ]^{6}
 ~\left [ \frac{E^*} {{\rm TeV}} \right ]^{-2}
\label{eq:nstarpnum}
\end{equation}
(where we have taken as ``standard reference'' the frequency of supernovae
in the Milky Way, and a high value for the critical energy $E^*$).
Observational evidence for the discreteness of CR protons will therefore
very likely be possible only via the observations of subtle, small effects.

For the case of large energy losses,
relevant for $e^\mp$ the critical energy $E^\dagger$ is in the other hand
many orders of magnitude smaller.
For example, for $\delta = 0.4$ (and therefore $\alpha = 2.4$)
one finds: 
\begin{equation}
 \left ( E^\dagger \right )_{e^\mp}
 \simeq 1.57 ~{\rm TeV}
 ~\left [ \frac{T_s} {50 ~{\rm yr}} \right ]^{-0.625}
 ~\left [ \frac{H} {5 ~{\rm kpc}} \right ]^{1.25}
 ~\left [ \frac{E^*} {3 ~{\rm GeV}} \right ]^{0.375} ~.
\label{eq:nstar2a}
\end{equation}
For $\delta = 0$ (that is rigidity independent diffusion), one has:
\begin{equation}
 \left ( E^\dagger \right )_{e^\mp}
 \simeq 0.490
~{\rm TeV}
 ~\left [ \frac{T_s} {50 ~{\rm yr}} \right ]^{-0.5}
 ~\left [ \frac{H} {5 ~{\rm kpc}} \right ]^{1.0}
 ~\left [ \frac{E^*} {3 ~{\rm GeV}} \right ]^{0.5}
~.
 \label{eq:nstar2b}
\end{equation}

In these estimates we have used as a reference low value of the
critical energy $E^*$, and the result is that in this case,
if the sources of the electrons are supernova explosions,\
then one expects very large effects for $E$ of order 1~TeV of less.
At the moment however there has been no clear evidence for effects
associated to the discreteness of the sources up to an energy of
order 10~TeV. To reconcile this absence of sourece discreteness effects with
the estimates given above one can assume that the source events
are significantly more frequent than a few per century.
An alternative solution is to assume that the critical energy
$E^*$ is larger, of order 1~TeV, so that CR propagate faster and more sources
can contribute to the flux.

\section{Interpretation}
\label{sec:interpretation}

In this work we have been arguing that two crucial problems
for the interpretation of the electron and positron cosmic rays 
are the identification of the effects of energy losses
and of the discreteness of the sources in the observed spectra.
After reviewing the existing observations in Sec.~\ref{sec:observations}
and theoretical models in Secs.~\ref{sec:models} and~\ref{sec:discrete}
it is now possible to summarize the main results.

\subsection{The critical energy $E^*$} 
\label{sec:ecrit}
On the basis of very general considerations one can predict that the spectra
of both electrons and positrons should exhibit a softening feature
at approximately the same energy $E^*$. At this energy
the residence time and energy loss time are approximately equal,
and the effects of energy loss become significant.
The shape and structure of these softening features
should be considered as model dependent,
but the existence of the features is a robust prediction.

Since the loss time $T_{\rm loss} (E)$ is determined by well known physics
(with the main uncertainty associated to the shape and size
of the CR confinement volume), the identification of $E^*$ 
corresponds to a measurement of the CR residence time at the energy $E^*$.

The phenomenological study of the $e^\mp$ spectra shows the existence of
one clear softening, observed for the sum $(e^- +e^+$), at $E \approx 1$~TeV.
At lower energy (where separate measurements of $e^-$ and $e^+$ spectra
are available) it is not easy to identify
spectral features that can be interpreted as the signature of energy
loss effects. A possibility however is to
place $E^*$ below 10~GeV, in the energy range
where the exponents of both spectra
change continuosly, and where solar modulations are important.

One can therefore conclude that there are
two alternative possibilities for the value of the critical energy $E^*$.
The first one is to have $E^* \lesssim 3$~GeV, 
well inside the region where solar modulations are important.
The second one is to identify the spectral break observed
in the $(e^-+e^+$) spectrum as the energy loss feature,
so that $E^* \simeq 1$~TeV.

The two solutions differ by a very large factor,
and therefore imply CR residence time that are very different.
A critical energy of order $E^* \simeq 3$~GeV implies
(for CR of the same rigidity) a residence time of order 200~Myr.
Such a long time is in tension
(if not in open conflict) with estimates of the CR residence time
based on measurements of the abundances of Beryllium isotopes \cite{beryllium}.

A critical energy of order $E^* \simeq 1$~TeV implies
a residence time of order 0.5~Myr.
The extrapolation at lower energy depends on the exponent $\delta$
of the escape time [since $T_{\rm esc} (E) = T_{\rm loss} (E^*) ~(E/E^*)^{-\delta}$].
For a rigidity of 10~GV one obtains a residence time
of order 1--4~Myr for $\delta = 0.1$--0.4.
This short residence time is not inconsistent with the Beryllium
isotope measurements, but it is in conflict with the grammage
estimated from the measurements of secondary nuclei
(Lithium, Beryllium and Boron), assuming that this grammage is accumulated
during propagation in interstellar space
\cite{Cowsik:2010zz,Cowsik:2013woa}.

To establish the validity of the low energy solution
for $E^*$ one needs a good understanding of the $e^\mp$ interstellar spectra.
These spectra are not directly observable, and must be
inferred from the measured ones correcting for 
the effects of solar modulations.
This procedure is at the moment model dependent.
In Sec.~\ref{sec:observations} we have shown that 
if the solar modulations are described with the Force Field
approximation the interstellar spectra of both
$e^-$ and $e^+$ can be well represented as unbroken power law
spectra in the energy range 1--10~GeV. This result is inconsistent
with the presence of a softening feature in the same energy range, and therefore
is in conflict with hypothesis of a low value for the critical energy.
More detailed studies of the
solar modulation effects could in the future
demonstrate that the interstellar $e^-$ and $e^+$ spectra deviate from a simple
smooth form the 1--10~GeV range, and that the
reconstructed spectral features
can be understood as the consequence of energy lossese.
Alternatively, these studies, could yield smooth interstellar spectra, 
and therefore exclude the possibility that
$E^*$ is at low energy.

The existence of the
spectral break at $E \approx 1$~TeV is clear,
however its shape appears narrower that what is predicted
by the models we have discussed in this work.
If this spectral structure is in fact 
generated by energy loss effects this could require improved models
for propagation. It is however also clear that it is very
desirable to obtain more precise measurements of the spectral shape
in this energy range.

It should also be added that if energy losses are not the explanation
for the spectral break at 1~TeV, one does need an alternative explanation
for its origin, and the apparent narrowness of the observed spectral break
is a problem also for these models.

\subsection{Source Granularity}
\label{sec:stochastic}

A second, robust prediction for the spectra of electrons, is that
is the particles are generated in rare events such as supernova explosions
or GRB's, the granularity of the source should become
visible at sufficiently high energy, when the number of source events
that contributes to the observed flux becomes small. 
The observable effects should be anisotropies for the angular distributions, and
deviations from a power law shape for tge energy spectra.
At very large $E$,
when the maximum propagation distance becomes shorter than
the distance of the closest source active in a time interval
of order $T_{\rm loss} (E)$,
the CR flux should become exponentially suppressed.

At the moment there are only upper limits \cite{Abdollahi:2017kyf}
on the anisotropy of the electron and positron spectra, and the measurements
above the spectral break at $E \simeq 1$~TeV
of DAMPE, HESS and the other Cherenkov telescopes are
consistent with a simple power law up to an energy of order 10--20~TeV.

These results are in tension with the hypothesis that the sources
of the electrons are supernova (SN) explosions (that have a known frequency
and space density), if the ``SN paradigm'' is combined with the assumption 
that cosmic rays have a long Galactic residence time
(and the critical energy $E^*$ is of order few GeV), because in this case
[see Eqs.~(\ref{eq:nstar2a}) and~(\ref{eq:nstar2b})] one expects large
effects already at $E \simeq 500$~GeV.

One explanation for this problem is to
deduce that the sources that generate the electrons are not
SN explosions, and have a higher frequency in the the Galaxy.
An alternative possibility is to conclude that
propagation distance for CR electrons (and positrons) is longer,
(and correspondingly the critical energy $E^*$ is higher)
so that more objects can contribute to the flux.

\section{Astrophysical implications}
\label{sec:astro}

\begin{figure}[bt]
\begin{center}
\includegraphics[width=12.0cm]{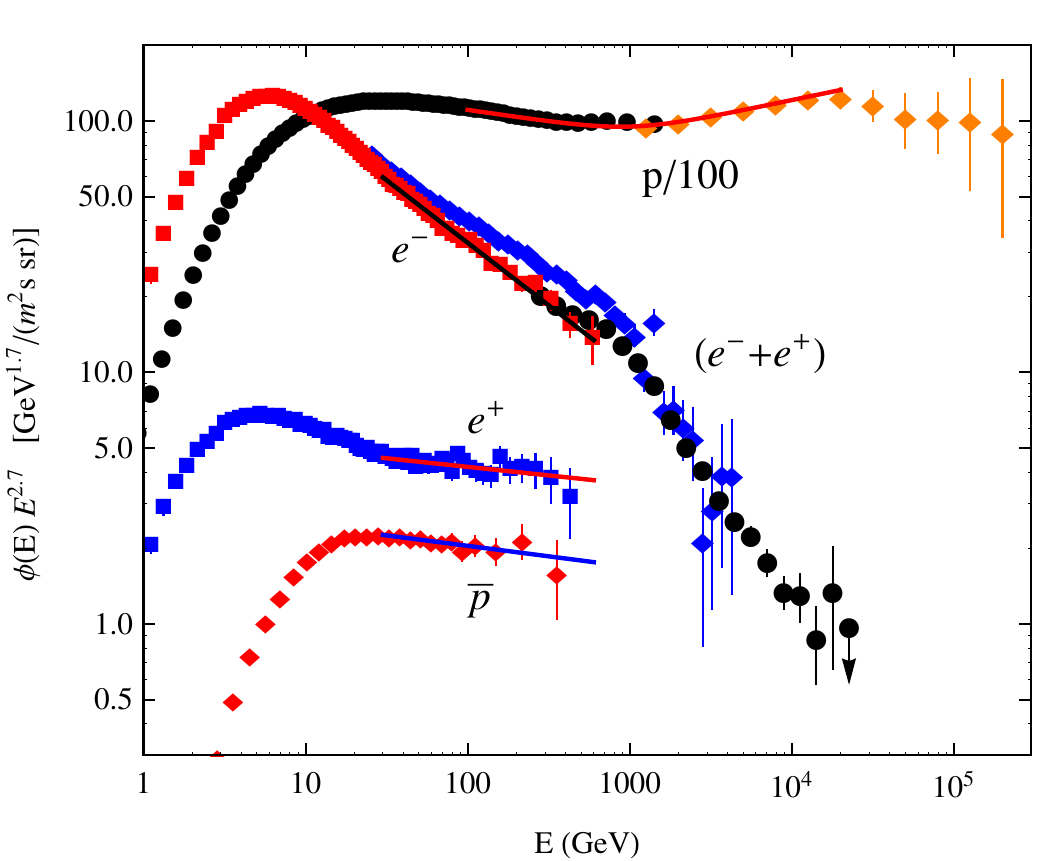}
\end{center}
\caption {\footnotesize
 Spectra of $p$, $e^-$, $e^+$, ($e^- + e^+$) and $\overline{p}$.
 The high energy data points for protons are from CREAM
 \protect\cite{Yoon:2017qjx}, the data points for
 ($e^- + e^+$) are from DAMPE and HESS. All other points are from AMS02.
 The lines superimposed to the $e^-$, $e^+$ and $\overline{p}$ data points
 are simple power law fits for $E > 30$~GeV. The line for the proton data is
 a broken power law fit taken from
\label{fig:fourfits} }
\end{figure}

The identification of the critical energy $E^*$ where energy losses
become important for the propagation in the Galaxy
of electron and positrons has profound implications for cosmic ray astrophysics.
These implications are particularly relevant for two problems:
\begin{enumerate}
 \item 
The relation between the acceleration of protons (and nuclei) versus
electrons in the CR sources.
\item
The possible existence of ``non standard'' mechanisms
for the generation of relativistic anti--particles
($e^+$ and $\overline{p}$) in addition to
the standard one, that is their production as secondaries in the
inelastic collisions of CR protons and nuclei.
\end{enumerate}
To illustrate the basic elements of these problems
one can study the spectra of 
$p$, $e^-$, $\overline{p}$ and $e^+$.
The four energy distributions are shown in Fig.~\ref{fig:fourfits}.
In the energy interval 30--400~GeV, the spectra of $e^\mp$ and $\overline{p}$ are
can be well described as simple power laws:
\begin{equation}
\phi_j (E) \simeq K_j ~\left ( \frac{E}{E_0} \right )^{-\gamma_j}~.
\label{eq:flux0}
\end{equation}
(with $E_0$ an arbitrary energy scale that we will 
fix at the value $E_0 = 50$~GeV).
The results of fits based in Eq.~(\ref{eq:flux0}) are
are listed in Table~\ref{tab:fit5}).

In the case of protons, an unbroken power law cannot describe the
spectrum because of the existence of a hardening around
$E \simeq 300$~GeV. The expression of Eq.~(\ref{eq:flux0}) can however
describe the $p$ spectrum in separate energy intervals.
For $E \lesssim 300$ ~GeV
the spectral index is of order $\gamma_p^{\rm low} \simeq 2.80$,
while at high energy it is of order $\gamma_p^{\rm high} \simeq 2.61$.
This high energy result is determined by the measurements of
CREAM \cite{Yoon:2017qjx}.

\begin{table}
 \caption{\footnotesize Power law fits to the CR spectra.
 \label{tab:fit5}}
\begin{center}
 \renewcommand{\arraystretch}{1.65}
 \begin{tabular}{ | l || c | c | c | c |}
\hline
Particle & $K$ (m$^2$~s~sr)$^{-1}$ & $\gamma$ & $N_{\rm d.o.f}$ & $\chi^2_{\rm min}$ \\
\hline 
$e^-$ & $(1.19 \pm 0.01) \times 10^{-3}$ & $3.20 \pm 0.01$ & 30 & 12.7 \\
$e^+$ & $(1.14 \pm 0.02) \times 10^{-4}$ & $2.77 \pm 0.02$ & 29 & 12.0 \\
$\overline{p}$ & $(5.6 \pm 0.2) \times 10^{-5}$ & $2.78 \pm 0.03$ & 12 & 1.6 \\
$p$ ~(low $E$) & $(0.301 \pm 0.002$) & $2.790 \pm 0.005$ & 12 & 1.6 \\
$p$ ~(high $E$) & $(0.210 \pm 0.001$) & $2.61 \pm 0.01$ & 9 & 19.7 \\
\hline
\end{tabular}
\end{center}
\end{table}

The comparision of the spectral shapes of the different particle types
shows some intriguing results:
\begin{itemize}
\item [(a)] The shapes of the $e^-$ and $p$ spectra are very different from each other,
 and the energy distributions of electrons is much softer.
 In the energy range 30--400 GeV $\gamma_p \simeq 2.80$, while
 $\gamma_{e^-} \simeq 3.20$. 
\item [(a)] The spectral indices of the $e^+$ and $\overline{p}$ fluxes
 are consistent with being equal to each other with value
 $\gamma_{e^+} \simeq \gamma_{\overline{p}} \simeq 2.78$.
 The ratio positron/antiproton 
 is approximately constant in the energy interval
 considered with value $e^+/\overline{p} \approx 2.02$.
\end{itemize}

The observation that the CR flux (for a certain particle type)
has a power law form in the energy interval
[$E_{\rm min},E_{\rm max}$] suggests that both the source
spectrum and the escape time $T_{\rm esc} (E)$ are of power law form
in the same interval. In the cases of electrons and positrons this also
imply (as already discussed) that the critical energy is outside the
energy interval considered, that is either below $E_{\rm min}$
(``low critical energy hypothesis'') or above $E_{\rm max}$ 
(``high critical energy hypothesis'').

If the standard mechanism of secondary production is the dominant source
of CR positrons and antiprotons, the source spectrum for
$E \gtrsim 10$~GeV (when threshold and mass effects are negligible)
is in fact in reasonably good approximation, of power law form,
with an exponent equal to the spectral index of the proton flux:
\begin{equation}
 q_{\overline{p} (e^+)} (E, \vec{x})\simeq {4 \, \pi} 
 ~ \sigma_{pp} \; \Sigma_{\rm ism} \; C_{\rm nuclei}
 ~Z_{pp \to \overline{p} (e^+)} (\gamma_p) ~K_p ~\left ( \frac{E}{E_0} \right )^{-\gamma_p}
 ~\delta [z]~.
\label{eq:q-secondaries}
\end{equation}
Writing this equation we have used the fact that interstellar gas in
concentrated in the Galactic disk,
$\sigma_{pp}$ is the inelastic cross section for $pp$ interactions,
$\Sigma_{\rm ism}$ is the density per unit area
of gas calculated integrating the density
along the $z$--axis, 
$C_{\rm nuclei} \simeq 1.7$ is an adimensional factor
that takes into account the contribution of other CR and target nuclei to the production
of antiparticles, and $Z_{pp \to j} (\gamma)$ is the so called ``$Z$--factor'', that
is the $(\gamma -1)$ moment
of the the inclusive spectrum for the production of particle $j$
in $pp$ collisions:
\begin{equation}
Z_{pp \to j} (\gamma) = \int_0^1 dx ~x^{\gamma -1} ~\frac{dn_{pp\to j} (x)}{dx} ~.
\end{equation}
Secondary particles of energy $E$ are generated by the collisions of protons
in a broad energy range $E_p \simeq 10$--100~$E$, and therefore
in the expression above one should use the parameters $K_p$ and $\gamma_p$
of protons at higher energy.

A very important point is that the source spectra of
antiprotons and positrons are proportional to each other
with an energy independent ratio:
\begin{equation}
 \frac{q_{e^+} (E)} {q_{\overline{p}} (E)} \simeq 
 \frac{Z_{pp \to e^+} (\gamma_p)} {Z_{pp \to \overline{p}}(\gamma_p) } \simeq 1.9 \pm 0.3 ~.
\end{equation}
The numerical estimate of this ratio (and its uncertainty) are determined
by the modeling of the properties of hadronic interactions
(see discussion in the appendix of \cite{Lipari:2016vqk}).

The observable CR fluxes can be obtained from the source spectra
using the results presented in Sec.~\ref{sec:diffusion}.
If energy losses are negligible one has:
\begin{equation}
 \phi_j (E) = \frac{c}{4 \, \pi} ~q_j (E) ~\frac{T_{\rm esc} (E)}{H}
 = \frac{c}{4 \, \pi} ~\frac{q_{0,j} \; T_{\rm esc} (E_0)}{H}
 ~ \left (\frac{E}{E_0} \right )^{-(\alpha_j + \delta)} ~,
 \label{eq:noloss}
\end{equation}
where $\alpha_j$ is the exponent of the source spectrum for particle type $j$,
and $\delta$ is the exponents that controls the rigidity dependence of the
diffusion time for all particles.

If energy losses are the main effect in the propagation,
then the CR flux is: 
\begin{equation}
 \phi_{j} (E)
 = 
 \frac{c}{4 \, \pi} ~\frac{q_{0,j} \; T_{\rm esc} (E_0)}{H}
 ~k(\alpha_j, \delta) \; \sqrt{1-\delta} 
 ~ \left (\frac{E^*}{E_0} \right )^{\frac{(1-\delta)}{2}} ~
 ~ \left (\frac{E}{E_0} \right )^{\alpha_j + \frac{(1+\delta)}{2}} ~.
 \label{eq:loss}
\end{equation}
with $k (\alpha, \delta)$ the constant of Eq.~(\ref{eq:kalpha}).

The important point of this discussion is that the spectral index $\gamma_j$
of the CR flux for particle $j$
is determined by the spectral index $\alpha_j$ of the source spectrum and
by the exponent $\delta$ of the diffusion time,
however the result depends in whether energy losses
are negiglible or not.
In the first case one has
\begin{equation}
\gamma_j = \alpha_j + \delta ~,
\label{eq:gamma1}
\end{equation}
in the second one:
\begin{equation}
 \gamma_j = \alpha_j + \frac{(1+\delta)}{2} ~.
\label{eq:gamma2}
\end{equation}
It is therefore clear that scenarios where the critical energy is small
($E^* < E_{\rm min}$)
or large
($E^* > E_{\rm max}$)
have profoundly different implications.

\subsection{High critical energy hypothesis}
If the critical energy $E^*$ is large
(above the maximum energy of the energy interval we are considering),
then the propagation properties of protons and electrons is approximately equal.
The observation that the spectral indices of $p$ and $e^-$ are
different, immediately implies that the two particles 
must have source spectra of different shape, and using Eq.~(\ref{eq:gamma1}) one
can deduce that $\alpha_{e^-} - \alpha_p \simeq \gamma_{e^-} - \gamma_p$.
Such a conclusion would obviously have important implications for the properties
of the CR accelerators.

On the other hand the assumptions that positrons and antiprotons
(in the energy interval considered) propagate in appoximately the same way
hypothesis is consistent with the hypothesis that the antiparticles fluxes
are generated by the standard mechanism. This consistency is quite striking because
it emerges from two independent observations. The first one is that the
spectral indices are equal within errors:
\begin{equation}
\gamma_{\overline{p}} \simeq \gamma_{e^+}
\label{eq:coinc1}
\end{equation}
The second one is that the ratio $e^+/\overline{p}$ is (within systematic uncertainties)
equal to the ratio for the source spectra:
\begin{equation}
 \frac{e^+} {\overline{p}}
 \equiv \frac{\phi_{e^+} (E)} {\phi_{\overline{p}}(E)} \simeq
 \frac{q_{e^+} (E)} {q_{\overline{p}}(E)} \simeq
 \frac{Z_{pp \to e^+} (\gamma_p)} {Z_{pp \to \overline{p}}(\gamma_p) } \approx 1.9 \pm 0.3 ~.
\label{eq:coinc2}
\end{equation}

Accepting the results that the antiparticle fluxes are generated
by the standard mechanism
(and therefore that $\alpha_{e^+} \simeq \alpha_{\overline{p}} \simeq \gamma_p$)
it becomes possible to estimate
the value of $\delta$ from Eq.~(\ref{eq:gamma1}) (the exponent that controls
the rigidity dependence of the diffusion), and also the spectral index of the
proton source $\alpha+p$:
\begin{equation}
 \delta
 \simeq \gamma_{\overline{p}} - \gamma_p 
 \simeq \gamma_{e^+} - \gamma_p \approx 0.2 ~
\end{equation}
and
\begin{equation}
 \alpha_p \simeq 
 \gamma_{p} - \delta \approx 2.4
\end{equation}
(where we have used for $\gamma_p$ the result of the fit of the
proton flux at high energy).
Combining Eqs.~(\ref{eq:noloss}) and~(\ref{eq:q-secondaries}) it is also
possible to estimate the quantity $T_{\rm esc}(E_0)/H \sim 1$~Myr/Kpc.

\subsection{Low critical energy hypothesis}
If the critical energy $E^*$ is smaller than $E_{\rm min}$, the relation between the
observed flux and the source spectrum is different for $p$ and $\overline{p}$
compared to $e^-$ and $e^+$. This offers the interesting possibility to
assume that the difference in shape for the fluxes of electrons and protons
is generated by propagation. Assuming $\alpha_p \simeq \alpha_{e^-}$,
one
can use Eq.~(\ref{eq:loss}) to estimate $\delta$, and the Eq.~(\ref{eq:noloss})
to estimate $\alpha_p$ with the results:
\begin{equation}
\delta = 1 - 2 \, (\gamma_{e^-} -\gamma_p ) \simeq 0.2 
\end{equation}
and (for $E \le 300$~GeV):
\begin{equation}
\alpha_p \simeq \gamma_p - \delta \simeq 2.6 
\end{equation}
Using Eq.~(\ref{eq:loss}) one can then estimate the
ratio of the electron and proton source spectra:
\begin{equation}
 \frac{q_{0,e}} {q_{0,p}} =
 \frac{K_e} {K_p}
 ~\left ( \frac{E^*}{E_0} \right )^{(\delta-1)/2}
 ~\frac{1}{k(\alpha, \delta)}
 \simeq 0.020 ~\left ( \frac{E^*}{3~{\rm GeV}} \right )^{-0.4} ~.
\end{equation}

The assumption that the propagation of $e^+$ and $\overline{p}$ is different
(because of energy loss effects for positrons) is however in clear conflict
with the hypothesis that the main source of antiparticles is secondary production.
This conflict is manifest in the observation that the
the spectral indices of the positron and antiproton fluxes are approximately equal.
The relation between the flux and the source spectral indices 
is now given by Eq.~(\ref{eq:gamma1})
for antiprotons, and by Eq.~(\ref{eq:gamma2}) for positrons,
so in this case the observation that $\gamma_{e^+} \simeq \gamma_{\overline{p}}$
implies that the shape of the source spectra are different and:
\begin{equation}
\alpha_{\overline{p}} - \alpha_{e^+} \simeq \frac{1-\delta}{2} ~.
\label{eq:coinc1a}
\end{equation}
This result is in conflict with the hypothesis that antiproton and positrons are
generated by the secondary production mechanism, or also by other
mechanisms (such as most models for dark matter self--annihilation or decay)
where the $e^+$ and $\overline{p}$ are generated with spectra of similar shape.

The bottom line is that the assumption that
the critical energy $E^*$ is below 10~GeV, requires the existence
of a new source of high energy positrons.
The two observations [see Eqs~\ref{eq:coinc1}) and~(\ref{eq:coinc2})]
that the spectral indices of the antiproton and positron fluxes
are approximately equal, and that the ratio $e^+/\overline{p}$ of the fluxes
is of order unity (and equal to the ratio of the hadronic $Z$ factors),
are now simply meaningless numerical ``coincidences'',
that constrain the shape and absolute normalization of the new positron source.

The conclusion that the positron flux contains a hard,
non standard component generates the crucial prediction that the spectrum
should exhibit a {\rm hardening} features associated to
the transition from the regime were the standard mechanism of secondary production
is the dominant mechanism of positron production to the regime
where the new source is dominant.
A discussion of the possible identification of this hardening
feature in the positron flux is contained in appendix~\ref{sec:hardening}.

\section{Outlook}
\label{sec:outlook}

The spectra of Galactic cosmic rays observable at the Earth
carry very important information
about their sources and about the structure of 
the magnetic field in our Galaxy.
The nuclear component (protons and nuclei)
dominates the flux, but the study of the smaller electron and positron components
is of great importance to understand the mechanisms that shape the spectra.
In some sense, the fluxes of $e^-$ and $e^+$ carry more information
than the fluxes of protons and nuclei
because their spectral shape should show the imprints of energy losses.

In this work we have argued that the effects of energy loss should
be significant only above a critical energy $E^*$, and very quickly
become dominant for propagation. The spectra of both $e^-$ and $e^+$ should
therefore exhibit softening features at approximately
the same energy that mark the onset of the regime in propagation
where energy losses are important.
The identification of the critical energy is a crucial problem
for CR astrophysics.

The study of the spectral shapes of of the
electron and positron spectrs suggest for $E^*$
two possible solutions that differ by a large factor,
and have profoundly different implications.

The first solution,
that is fact implicitely the ``orthodody'' most commonly
(but not universally) accepted is
that the critical energy has a value of order few GeV.
This implies a very long residence time for CR in the Galaxy,
and a non--standard hard source for positrons in the Galaxy.
This ``low critical energy'' scenario is consistent with the hypothesis
that the CR accelerators release
in interstellar space spectra of $e^-$ and $p$ that,
in a broad energy range have the same shape.
In this scenario, the softening features associated to the energy loss
are in an energy range where the CR spectra are distorted by
solar modulations. Understanding the (interstellar) shape of the spectra
requires therefore a sufficiently good understanding of the solar modulations effects.
In the present work we have shown using a very simple models
for the modulations, that the observations are possibly consistent
with interstellar spectra that in the regime $E \lesssim 30$~GeV are unbroken
power laws, in conflict with the hypothesis that the critical energy
$E^*$ is in this energy range. Additional studies, based on more
detailed modelings of solar modulations are
clearly needed to reach a more firm conclusion.

The alternative solution for the critical energy $E^*$
is to identify the softening associated to energy loss effects with
the spectral break observed in the ($e^- + e^+$) spectrum at
$E \simeq 1$~TeV. This solution implies a much shorter CR residence time
in the Galaxy, probably in conflict
with models where the source of secondary nuclei are nuclear
fragmentation in interstellar space.

An attractive consequence of this `` high critical
energy scenario'' is that it provides a
natural explanation for the intriguing result
that the spectra of positrons and antiprotons have (in a broad energy interval from
30--400~GeV approximately the same spectral index and a ratio
$e^+/\overline{p} \simeq 2$,
simply assuming that the dominant source for both antiparticles is secondary
production. This is a straightforward consequence of the fact that
in this energy range the propagation of positrons and antiprotons is approximately
equal because the energy losses are small for both particles
(even if the rates of energy loss differ by many orders of magnitude).
An additional attractive feature of this solution for the critical energy
is that it provides a natural, simple interpretation for the TeV break
in the all--electron spectrum as the imprint of energy losses.
If $E^*$ is below 10~GeV it is necessary to
postulate a different mechanism as the origin of the spectral break
around 1~TeV, and this is not a trivial task.
The apparent narrowness of the observed spectral feature is a
problem for all models that address the origin of this structure.

An important problem for the ``high critical energy scenario''
is that it requires source spectra of electrons and protons
with very different shape. If this is correct the consequences for
the properties of the CR sources are clearly very important.

An additional prediction for the electron spectrum
(and possibly also for the positron spectrum) is that
at sufficiently high energy (when energy losses are important) the space--time
volume of the source shrinks rapidly with $E$, and is (as predicted)
the sources are point--like, transient astrophysical objects,
the granularity of the sources should leave observable signature for
the angular and energy distributions.
The minimum energy where the ``source granularity effects'' should become observable
grows with $E^*$.
The non--observation of source granularity
effects, and the measurement for $(e^- + e^+)$ spectrum of a shape
consistent with a simple power law above the break at 1~TeV
(up to the highest energy where measurements are available),
appears to be in serious tension (of not open confict) with the hypothesis
that the source of CR electrons are SN explosions, if one choses for $E^*$
the low energy solution. The high critical energy scenario is consistenty
with the idea that SN are the source of electrons, but the evidence for the
discreteness of the source should become observable soon.

The main conclusions of this discussion is that the question
of when energy loss effects become important for the propagation of
electrons and positrons is of central importance for CR studies.
Intimately connected to this question are the problems of 
understanding the physical mechanism that generates
the break at TeV energy in ($e^- + e^+$) spectrum, and the search for
``source granularity effects'' in the spectra of electrons and positron.'

Several lines of investigation, both theoretical and experimental
promise to make progress toward a clarification for these problems.
\begin{itemize}
\item Extend the measurements of the (separate) positron and electron spectra
 to higher energy.
 If the $e^-$ and $e^+$ spectra have a spectral breaks
 with a significantly different shape, the ``high critical energy scenarios''
 could be excluded (and in the opposite case reinforced).

\item Measure with precision the shape of the spectral break around one TeV.
 Understanding of the origin of the discrepancies
 between different measurements in this energy range is very desirable.

\item Extend the measurements of the ($e^- + e^+$) spectrum to as high energy as possible,
 with the hope to identify some of the sources. 

 \item Develop a better understanding of the solar modulation effects
 to confirm (or refute) the ``low critical energy scenario''.

 \item Study the spectra of electrons and positrons in different regions of the
 Galaxy (interpreting the measurements of the diffuse gamma ray flux).

 \item Construct better models for the propagation of
 CR in the Galaxy. 
\end{itemize}

\vspace{0.25 cm}

\noindent{\bf Acknowledgments.} During the preparation of this work I have benefitted from
discussion with several colleagues. I'm grateful to
Pasquale Blasi,
Carmelo Evoli,
Stefano Gabici,
Daniele Gaggero,
Tom Gaisser,
Dario Grasso,
Philipp Mertsch, 
Andrew Strong 
and
Elena Orlando.

\clearpage

\appendix

\section{Multiple components in the $e^\mp$ spectra}
\label{sec:hardening}
A large number of models for the interpretation of the
positron flux assume that the spectrum is formed by two components,
generated by different mechanisms,
and predict the existence of a hardening feature associated to 
the transition from a (low energy) regime dominated by one
component to a (high energy) regime dominated by the second component.

As discussed in the main texts, if one assumes that the critical energy
is $E^* \lesssim 10$~GeV, the observations require the
existence of a hard source of positrons, together with the standard mechanism
of secondary production in the inelastic interactions of protons and nuclei.

It is straightforward that in this situation it is natural
to expect that the spectrum contains a hardening feature.
In the simplest possible situation,
the two components can be described by power laws with different exponents:
\begin{equation}
 \phi(E) =
 K_1 \; \left ( \frac{E}{E_0} \right )^{-\gamma_1} 
+ K_2 \; \left ( \frac{E}{E_0} \right )^{-\gamma_2} 
\end{equation}
If $\gamma_1 > \gamma_2$, the first component has the softest spectrum,
and is dominant at low energy.
The cross--over energy where the two components are equal is:
\begin{equation}
E_{\rm cross} = E_0 \; \left (\frac{K_1}{K_2} \right )^{1/(\gamma_1 - \gamma_2)}~.
\end{equation}
and the spectrum can also be expressed in the form:
\begin{equation}
 \phi(E) = 
 K_1 \; \left ( \frac{E}{E_0} \right )^{-\gamma_1} \;
 \left [1 + \left ( \frac{E}{E_{\rm cross}} \right )^{-(\gamma_2 - \gamma_1)} \right ] ~
\label{eq:twocomponents}
\end{equation}
Comparing Eqs.~(\ref{eq:twocomponents})
and (\ref{eq:break-parametrization}) one can see that the two model components
is exactly described by the form
of Eq.~(\ref{eq:break-parametrization}), when the width parameter has the value:
\begin{equation}
w = \frac{1}{\gamma_1 - \gamma_2}~.
\label{eq:2-components}
\end{equation}
In fact, the widths of the best fits for positrons and electrons
estimated in Sec.~\ref{sec:observations} are of the same order of
$(\gamma_1-\gamma_2)^{-1}$, and 
therefore it is interesting to verify if it is possible to
describe the hardenings in the positron and electron spectra as the
manifestations of the the presence of two components.
In order to do this one can fit the observations using
as a template the expression in
Eq.~(\ref{eq:twocomponents}) distorted by solar modulations
(described by the FFA algorithm), that is with the 5 parameter form:
\begin{equation}
 \phi(E) = K_1 \;
\frac{E^2}{(E+ \varepsilon)^2} \;
\left ( \frac{E+\varepsilon}{E_0} \right )^{-\gamma_1} \;
 \left [1 + \left ( \frac{E}{E_{\rm cross}} \right )^{-(\gamma_2 - \gamma_1)} \right ] ~
\label{eq:twocomponents-mod}
\end{equation}
The results of fits performed on the basis of
Eq.~(\ref{eq:twocomponents-mod})
are given in table~\ref{tab:fit4}, and are of good quality.

\begin{table}
 \caption{\footnotesize Parameters of fits to the electron and positron spectra
 measured by AMS02.
 \label{tab:fit4}}
\begin{center}
 \renewcommand{\arraystretch}{1.65}
 \begin{tabular}{ | l || c | c |}
\hline
~~ & Electrons & Positrons \\
\hline 
$K$ [GeV\;m$^2$\;s\;sr]$^{-1}$ & $0.40\pm 0.02$ & $0.014 \pm 0.001$ \\
$\gamma_1$ & $4.26^{+0.15}_{-0.14}$ & $3.99^{+0.21}_{-0.18}$ \\
$\gamma_2$ & $3.07^{+0.06}_{-0.07}$ & $2.68^{+0.07}_{-0.08}$ \\
$E_{\rm cross}$ [GeV] & $24.2^{+7.3}_{-4.5}$ & $15.64^{+5.0}_{-2.9}$ \\
$\varepsilon$ [GeV] & $1.63\pm 0.10$ & $1.09^{+0.12}_{-0.11}$ \\
\hline
$\chi^2_{\rm min}$ & 14.1 & 21.6 \\
$N_{\rm d.o.f.}$ & $65$ & $64$ \\
\hline
\end{tabular}
\end{center}
\end{table}

In the case of positrons the fit has $\chi^2_{\rm min} = 21.6$
(for 64 d.o.f.) only marginally worse ($\Delta \chi^2 \simeq 0.6$)
than the more general fit performed with the model of Eq.~(\ref{eq:fit-form1}),
when in fact one has one less parameter.
For the electron flux one has $\chi^2_{\rm min} = 14.1$ (for 65 d.o.f.).
This has $\Delta \chi^2 \simeq 2.4$ with respect to the fit performed in the more
general model where the width $w$ is a free parameter.
The crossing energy for the best fits are
$E_{\rm cross} \simeq 24.2^{+7.3}_{-4.5}$ for electrons and 
$E_{\rm cross} \simeq 15.6^{+5.0}_{-2.9}$ for positrons. 

The crucial question is if the two components that emerge
from this fit procedure are real, or simply a mathematical artefact.

A reason to doubt that the hardening seen in the positron flux
is associated to a new component 
is the fact that hardenings of similar structure
are seen in both the $e^-$ and $e^+$ spectra.
Most models for a new source of positrons
(including acceleration from young Pulsars and creation
in dark matter particles self--annihilation or decay)
predict that the new mechanism generates
approximately equal source spectra of $e^-$ and $e^+$.

The hypothesis of a new hard component with equal
contributions to ther $e^-$ and $e^+$ spectra is however
not consistent with the data.
In the energy range 10--30~GeV, where the spectral hardenings
are visible, the electron flux is more than one order of magnitude
larger than the positron flux, and therefore the
second (hard) component for electrons is much larger that the same
component for positrons (see also Table~\ref{tab:fit4}).

The existence of the hardening in the $e^-$ and $e^+$ spectra remains
an intriguing feature, perhaps generated by propagation effects, that
has not yet found a convincing explanation.

\end{document}